\documentclass[11pt,a4paper]{article}

\pdfoutput=1

\usepackage{jheppub}
\usepackage{amssymb,amsmath,amsthm,graphicx,mathrsfs,bbm}
\usepackage{latexsym,amscd,amsbsy,amsfonts,dsfont}
\usepackage{graphics, color}
\usepackage{subfigure}
\usepackage{bm}
\usepackage{epsfig}
\usepackage{textcomp}
\usepackage[utf8]{inputenc}
\usepackage{graphicx}
\usepackage{xcolor}
\usepackage{multirow}
\usepackage{makecell}

\allowdisplaybreaks[1]

\makeatletter
\def\@fpheader{\relax}
\makeatother

\newcommand{\beq}{\begin{equation}}
\newcommand{\eeq}{\end{equation}}
\newcommand{\beqa}{\begin{eqnarray}}
\newcommand{\eeqa}{\end{eqnarray}}
\newcommand{\bea}{\begin{eqnarray}}
\newcommand{\eea}{\end{eqnarray}}

\newcommand{\lp}{\left(}
\newcommand{\rp}{\right)}

\newcommand{\ord}[1]{{\mathcal O}\lp #1\rp}

\begin{document}

\title{Holographic duals of evaporating black holes}
\author[1,2]{Roberto Emparan,}
\emailAdd{emparan@ub.edu}
\author[3]{Raimon Luna,}
\emailAdd{raimon.luna-perello@uv.es}
\author[4]{Ryotaku Suzuki,}
\emailAdd{sryotaku@toyota-ti.ac.jp}
\author[5]{Marija Tomašević,}
\emailAdd{marija.tomasevic@polytechnique.edu}
\author[2]{Benson~Way}
\emailAdd{benson@icc.ub.edu}

\affiliation[1]{Instituci\'o Catalana de Recerca i Estudis
Avan\c cats (ICREA)\\ Passeig Llu\'{\i}s Companys 23, E-08010 Barcelona,}
\affiliation[2]{Departament de F\'{i}sica Qu\`{a}ntica i Astrof\'{i}sica, Institut de Ci\`{e}ncies del Cosmos,\\
Universitat de Barcelona, Mart\'{i} i Franqu\`{e}s, 1, E-08028 Barcelona, Spain}
\affiliation[3]{Departament d'Astronomia i Astrof\'{\i}sica, Universitat de Val\`encia, Dr. Moliner 50, 46100, Burjassot, Val\`encia, Spain}
\affiliation[4]{Mathematical Physics Laboratory, Toyota Technological Institute,\\ Hisakata 2-12-1, Nagoya 468-8511, Japan.}
\affiliation[5]{CPHT, CNRS, École polytechnique, Institut Polytechnique de Paris, 91120 Palaiseau, France}

\newcommand{\blue}{\color{blue}}
\newcommand{\red}{\color{red}}

\abstract{
We describe the dynamical evaporation of a black hole as the classical evolution in time of a black hole in an Anti-de Sitter braneworld. A bulk black hole whose horizon intersects the brane yields the classical bulk dual of a black hole coupled to quantum conformal fields. The evaporation of this black hole happens when the bulk horizon slides off the brane, making the horizon on the brane shrink. We use a large-$D$ effective theory of the bulk Einstein equations to solve the time evolution of these systems. With this method, we study the dual evaporation of a variety of black holes interacting with colder radiation baths. We also obtain the dual of the collapse of holographic radiation to form a black hole on the brane. Finally, we discuss the evolution of the Page curve of the radiation in our evaporation setups, with entanglement islands appearing and then shrinking during the decreasing part of the curve. 
}

\begin{flushright}
CPHT-RR057.112022, TTI-MATHPHYS-18
\end{flushright}

\maketitle

\section{Introduction and Summary}

The AdS/CFT correspondence maps the problem of solving a quantum field theory in a background spacetime $\mathcal{B}_d$ onto the task of finding a regular solution of the Einstein equations in AdS$_{d+1}$ with the geometry of $\mathcal{B}_d$ at its conformal boundary. This AdS bulk holographically encodes the state of the quantum fields on the fixed, non-dynamical background $\mathcal{B}_d$. Then, if $\mathcal{B}_d$ is a black hole spacetime, we obtain a powerful approach to study the Hawking radiation of interacting quantum fields emitted by that black hole. Ref.~\cite{Marolf:2013ioa} reviews early applications of this method in several black hole backgrounds.

Even earlier, it was proposed in \cite{Tanaka:2002rb,Emparan:2002px} that holography can actually address the harder problem of the dynamical evolution of an evaporating black hole. To achieve this, one resorts to braneworld holography. If the boundary is not at asymptotic infinity but on a brane at a  finite distance in the bulk, then dynamical gravity is induced on $\mathcal{B}_d$. This brane naturally responds to fluctuations in the bulk geometry, which is itself the dual description of a (cutoff) CFT. As a result, solving the classical dynamics of this AdS braneworld is equivalent to solving the semiclassical Einstein equations in one less dimension,
\begin{equation}
    G_{ij}+\dots=8\pi G_d\langle T_{ij}\rangle\,.
\end{equation}
Here $G_{ij}$ is the Einstein tensor (plus possibly a cosmological constant term) for the metric on the brane, $\langle T_{ij}\rangle$ is the renormalized stress-energy tensor of the holographic CFT, and the dots indicate higher-curvature terms (generated by the integration of the CFT ultraviolet degrees of freedom above the cutoff imposed by the brane).~This approach naturally incorporates the backreaction of the quantum radiation on the classical black hole geometry, which led the authors of \cite{Tanaka:2002rb,Emparan:2002px} to conjecture that it should be possible to study the Hawking evaporation process by solving the Einstein equations for a black hole in an AdS braneworld.

\paragraph{Difficulties for holographic evaporation.} As it turns out, the properties of weakly interacting quantum fields in the presence of a black hole can be an unreliable guide to the behavior of holographic CFTs. Using Einstein's classical theory in the AdS bulk implies that the dual CFT is a large-$N$, strongly coupled quantum theory---strictly speaking, both $N$ and the coupling are infinite. Although quantum fields in the presence of black holes present thermal effects independently of their interactions, holographic CFTs can have unusual properties. In particular, there are static states of the CFT in thermal equilibrium with a black hole (that is, they satisfy the Kubo-Martin-Schwinger condition) which nevertheless do not have any thermal radiation near infinity. 

\begin{figure}
        \centering
         \includegraphics[width=.5\textwidth]{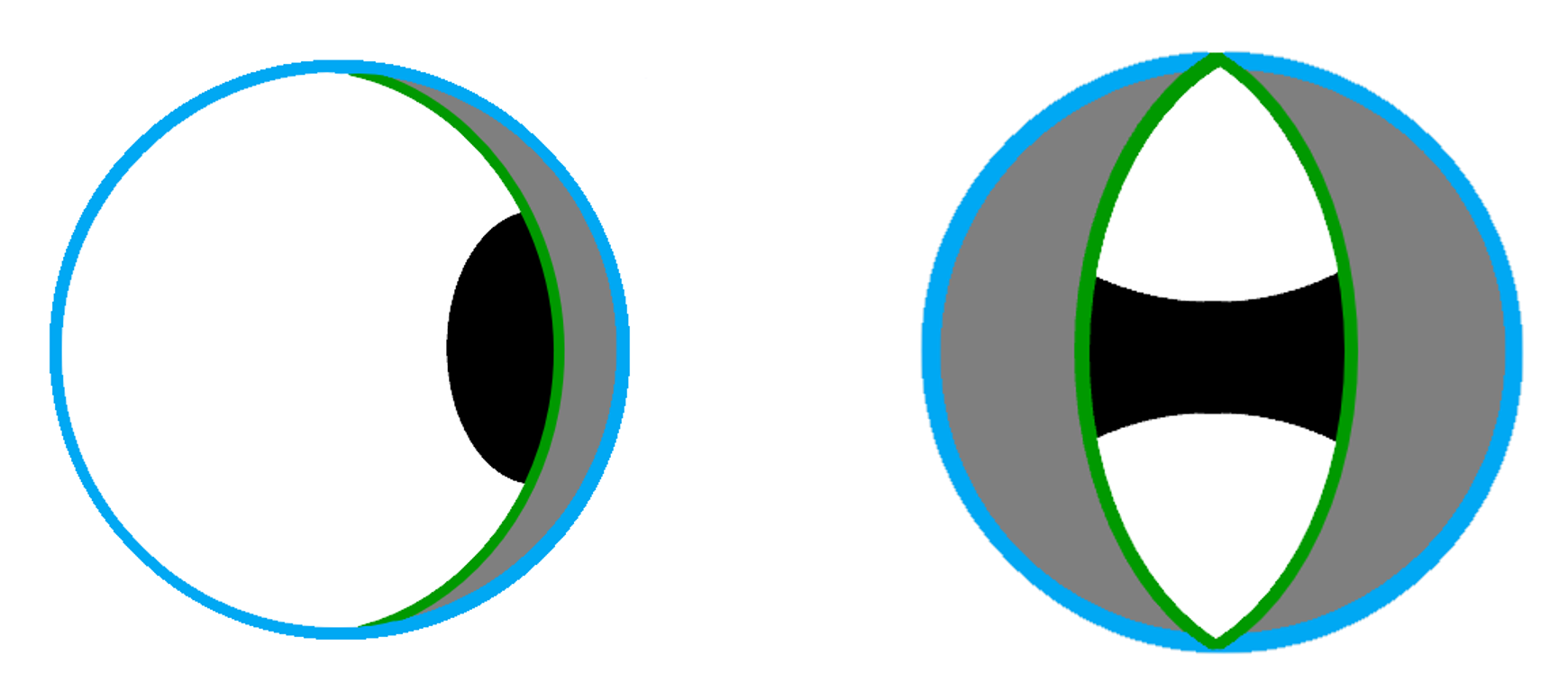}
        \caption{\small Two different bulk configurations for static brane black holes. The blue circle represents the asymptotic AdS boundary. In a braneworld construction, part or all of this boundary is absent by the introduction of (green) branes, which remove the gray-shaded portions of the bulk. Here we illustrate the configurations for branes with AdS geometry, but they also exist for flat (Randall-Sundrum) branes.       
        \textbf{Left:} Single black droplet. Its dual on the brane is a black hole surrounded by a CFT halo. \textbf{Right:} Black string, or uniform funnel. Its holographic CFT stress tensor is zero (other than possibly a Weyl anomaly). There also exists a state in which we have a twin droplet---a two-brane version of the single droplet---which represents a different state of the CFT coupled to the two black holes, with different stress tensor and different entanglement structure.}
        \label{fig:droplets and funnel}
\end{figure}

Two instances of this phenomenon are the following (see figure~\ref{fig:droplets and funnel}). In \cite{Figueras:2011va,Figueras:2011gd} a \textit{black droplet} in AdS$_5$ was numerically constructed whose dual is a four-dimensional Schwarzschild black hole surrounded by a static halo of polarized CFT,\footnote{See \cite{Tanahashi:2007wt,Kashiyama:2009pw,Abdolrahimi:2012qi,Abdolrahimi:2012pb,Banerjee:2021qei} for other approaches to this problem, and \cite{Figueras:2013jja,Biggs:2021iqw} for the addition of rotation.} resembling more a Casimir effect than Hawking emission since there is no radiation at a large distance. A simpler, and perhaps even more puzzling configuration is the AdS black string. Its dual describes a state of a CFT in the presence of two black holes at the antipodes of a static spherical Einstein universe. Not only is Hawking emission absent in this state: the quantum stress tensor is exactly zero everywhere (other than possibly for the Weyl anomaly) \cite{Gregory:2008br}. Although the leading $1/N^2$ corrections will yield a small, $O(N^0)$ radiation flux due to bulk quantum Hawking emission, the larger, $O(N^2)$ effect that we are interested in, which corresponds to classical bulk dynamics, is absent in these two examples.

These are configurations where hot black holes do not radiate to cold infinity. How is such perfect insulation possible? A plausible explanation attributes it to the infinite $N$ and infinitely strong coupling of the theory. 

The radiation from the black hole must, by gauge invariance, be color neutral. It could consist of ``glueballs",\footnote{By this we refer to the CFT duals of bulk gravitons, whether the CFT is in a confined state or not. As we will see, confinement at an infrared scale is not relevant to this discussion.} but this radiation would have $O(1)$ energy, and so would be invisible to the classical bulk geometry, which only encodes $O(N^2)$ energies. That is, the radiation of $O(N^2)$ degrees of freedom requires a classical dynamical gravitational configuration in the bulk, and furthermore, if these CFT degrees of freedom are in an approximately thermal state (such as a plasma, made of colored gluons and quarks but overall color-neutral), then the only possibility is that there is a black hole horizon moving away from the brane into the bulk.\footnote{The emission of a gravitational wave in the bulk would also be an $O(N^2)$ effect in the CFT, but a coherent one, far from thermality.} In order to incorporate a mechanism of this kind for the dual evaporation of a black droplet, Refs.~\cite{Tanaka:2002rb,Emparan:2002px} proposed that the latter suffers from a classical instability akin to the Gregory-Laflamme instability of black strings \cite{Gregory:1993vy}. Presumably, this would lead to the breakup of the droplet with (portions of) the horizon falling into the bulk, thus realizing the outgoing thermal radiation. But, as emphasized in \cite{Fitzpatrick:2006cd}, a black droplet is unlikely to spontaneously pinch off unless it is long enough to trigger the instability. 

Note that having $N\to\infty$ is not, by itself, enough to suppress the radiation: a black hole coupled to $N^2\gg 1$ free fields will radiate much like Hawking predicted, only $N^2$ times faster. Thus, a strong enough coupling appears to be necessary to quench the emission. Moreover, the presence or not of confinement at an infrared scale does not seem relevant to explain the existence of static, stable droplets. These exist both at the asymptotically flat boundary of Poincaré AdS, with a deconfined CFT, and on the spherical universe at the global AdS boundary, where the CFT has a gap at any $N$ and any coupling, exhibiting `kinematic confinement'. A classical black droplet at (or near) the boundary of AdS is decoupled from any infrared degrees of freedom of the CFT, and this causes a complete disconnection from any low-energy thermal radiation, regardless of whether the theory is gapped or not.\footnote{See \cite{Kaloper:2012hu} for further investigation of this problem.} 

Apparently, then, brane black holes remain stuck to the brane, so that, in dual terms, the evaporation of a black hole is thwarted in the holographic limit of infinite $N$ and infinite coupling. 

\paragraph{Holographic evaporation redux.} In this article we show that this conclusion is not the end of the story: The evaporation of a black hole as the dual process of a horizon sliding off the brane is possible, and is governed by physics akin to the Gregory-Laflamme instability (that is, dynamics of horizons with several separate length scales). We will solve the dynamical Einstein equations in the bulk in several setups where we explicitly demonstrate the phenomenon.

The main element enabling the dual evaporation is the connection of the brane black hole to at least some (approximately) thermal degrees of freedom of the CFT, i.e., dual to a bulk horizon, even if this is smaller than the AdS radius. 
This is achieved by connecting the black hole on the brane to another horizon in the bulk (often viewed in dual terms as a large enough ``bath'', though not necessarily of infinite size), which we will do in a variety of manners. Once all parts of the system are joined via a horizon in the bulk through which heat (i.e., horizon generators) can flow, we will see that the elementary principle that
\begin{quote}
\centering
\emph{heat (radiation) flows from the hotter regions into the colder ones}
\end{quote}
allows us to understand how all our examples work.\footnote{Incidentally, the principle that ``heat flows from the hotter parts of the horizon to the colder ones'' is also generally applicable to understanding the Gregory-Laflamme instability. This can be made precise and rigorous in the hydrodynamic limit of black strings and branes \cite{Bhattacharyya:2007vjd,Emparan:2009at}, but it is qualitatively valid more widely.\label{foot:GL}}

Of course, this simple idea acquires value only after we provide a convenient means to identify the hotter and colder parts of our systems. For this, an aspect of our study that was not present in the early investigations (at least not prominently so) will be important. Namely, we consider global AdS bulks and Karch-Randall branes with AdS geometry. Besides providing a natural infrared regulator for the CFT, global AdS allows for richer phases and dynamics than Poincar\'e-AdS constructions \cite{Hirayama:2001bi,Marolf:2019wkz,Emparan:2021ewh,Licht:2022rke}. In particular, black holes in global AdS with cosmological radius $L$ come in two main varieties, namely,
\begin{itemize}
    \item small AdS black holes, with horizon size $r_0<L$, which are colder the bigger they are;
    \item large AdS black holes, with $r_0>L$, which are hotter the bigger they are.
\end{itemize}

Another relevant feature of the global AdS setup is that sufficiently thin black strings in AdS, with $r_0\lesssim L$ (we will give precise values later), suffer from Gregory-Laflamme instabilities, while fatter ones are dynamically stable \cite{Hirayama:2001bi,Emparan:2021ewh}. The `planar black strings' obtained when $r_0/L\to\infty$ are always stable. Furthermore, we have verified that small AdS black droplets are dynamically stable and do not slide off the brane, supporting the view that the black droplets of \cite{Figueras:2011gd} are indeed stable.

The last new ingredient in our analysis is methodological and is crucial to efficiently tackle the temporal evolution of the classical bulk system. For this purpose, we resort to an effective theory of black holes in the limit of a large number of dimensions, $D\to\infty$ (see \cite{Bhattacharyya:2015dva,Bhattacharyya:2015fdk,Dandekar:2016jrp} for a parallel approach, and \cite{Emparan:2020inr} for a review). This method reduces the dynamics of the horizons to a set of two PDEs that can be quickly solved numerically in a conventional computer. Although the leading large-$D$ approximation may not give quantitatively accurate results for, say, AdS$_5$/CFT$_4$, it has nevertheless proven to be a fruitful guide to qualitatively unravel several difficult black hole problems \cite{Emparan:2015gva,Andrade:2019edf,Emparan:2021ewh,Licht:2022rke,Luna:2022tgh}.

\paragraph{Solving holographic collapse and evaporation.}

\begin{figure}
        \centering
         \includegraphics[width=.5\textwidth]{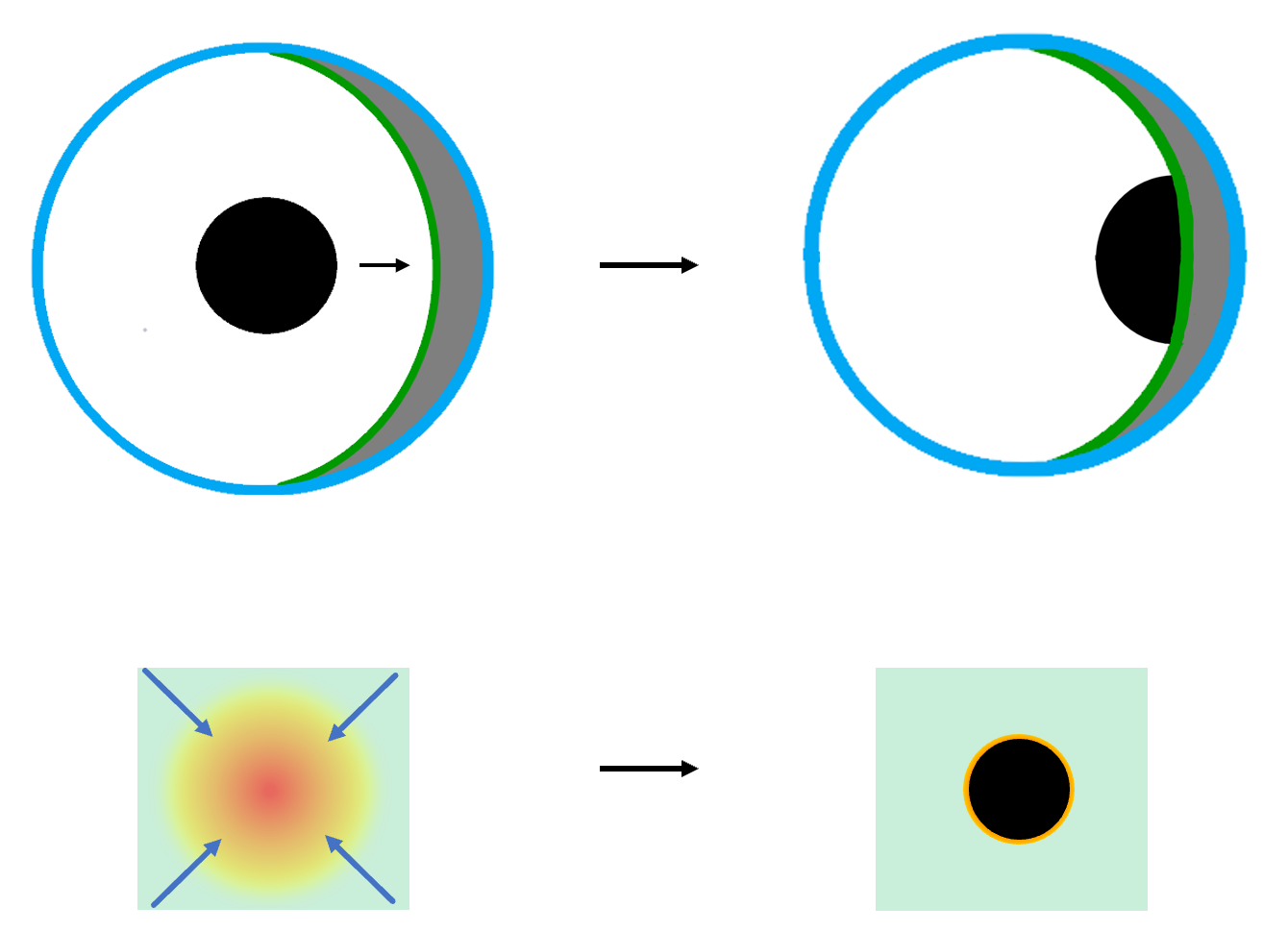}
        \caption{\small \textbf{Upper}: A bulk black hole is thrown towards the brane and sticks to it. \textbf{Lower:} dual view of the collapse of CFT radiation into a black hole. 
        In this and subsequent figures, we show first the bulk classical evolution, and then the dual view of black holes coupled to a holographic CFT.}
        \label{fig:collapsing}
\end{figure}

With this technique, we are able to holographically model the formation and evaporation of a brane black hole. For instance, we can model the dynamical collapse of a CFT cloud to form a black hole, by throwing a bulk black hole towards the brane. The black hole sticks to the brane, resulting in a stable black droplet; see figure~\ref{fig:collapsing}. 
Another model of collapse has a large, hot cloud of CFT falling onto a black hole and being absorbed by it. This is dually obtained with an initially static, large-AdS bulk black hole connected to the brane; when it moves towards the brane and sticks to it, it yields a long-lived black droplet.\footnote{A different kind of braneworld collapse to a black hole, using classical scalar fields on the brane, was studied in \cite{Wang:2016nqi}.}

We then describe different evaporation setups. In all of them, the horizon of a black hole that initially intersects the brane evolves to shrink in size on the brane, while the horizon generators flow into the bulk. The bulk evolution is classical, so even though the horizon on the brane reduces its size, in the bulk it does not become smaller; in fact, typically it grows. That is, the horizon slides off the brane into the bulk. This is the dual of the evaporation of a black hole into the (approximately) thermalized $O(N^2)$ degrees of freedom of the CFT.

\begin{figure}
        \centering
         \includegraphics[width=.9\textwidth]{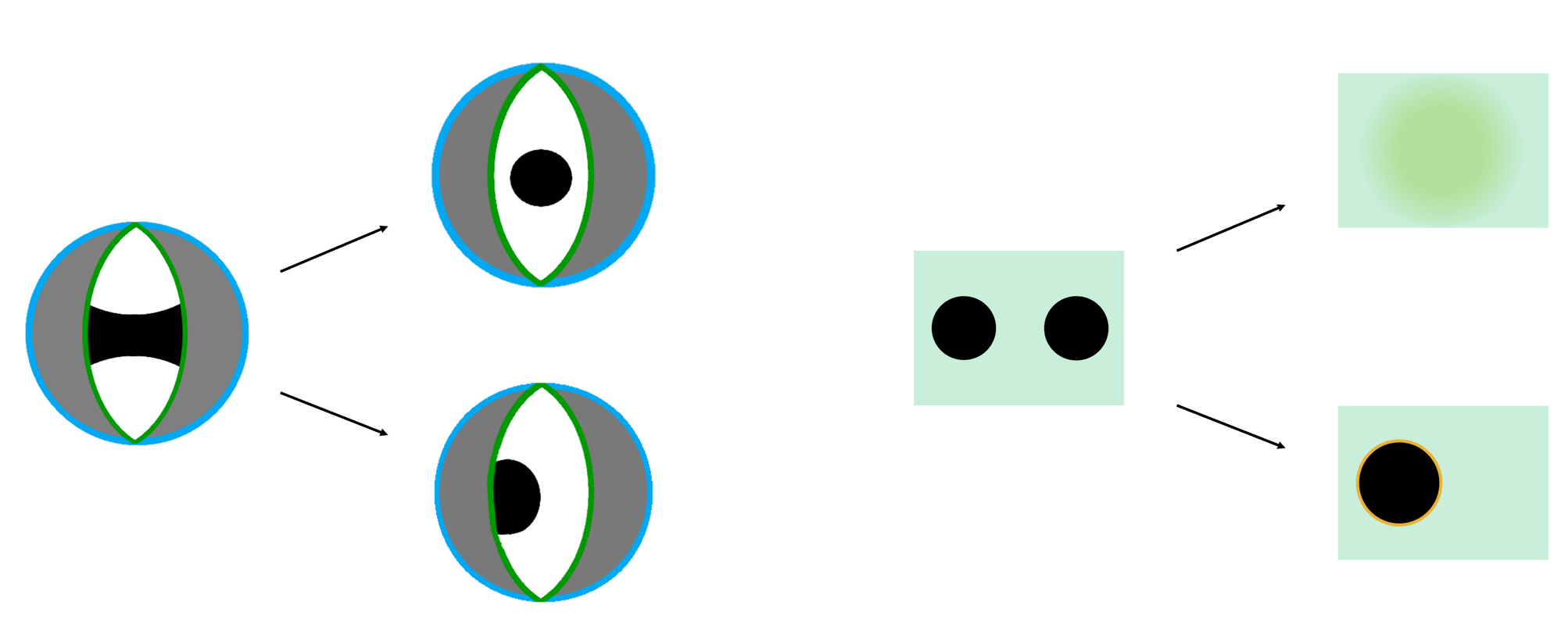}
        \caption{\small Two scenarios for funnel black hole evaporation, triggered by different perturbations. \textbf{Upper:} Both brane black holes evaporate into the bulk. \textbf{Lower:} One brane black hole evaporates into the other. }
        \label{fig:funneling}
\end{figure}

\begin{figure}
        \centering
         \includegraphics[width=.5\textwidth]{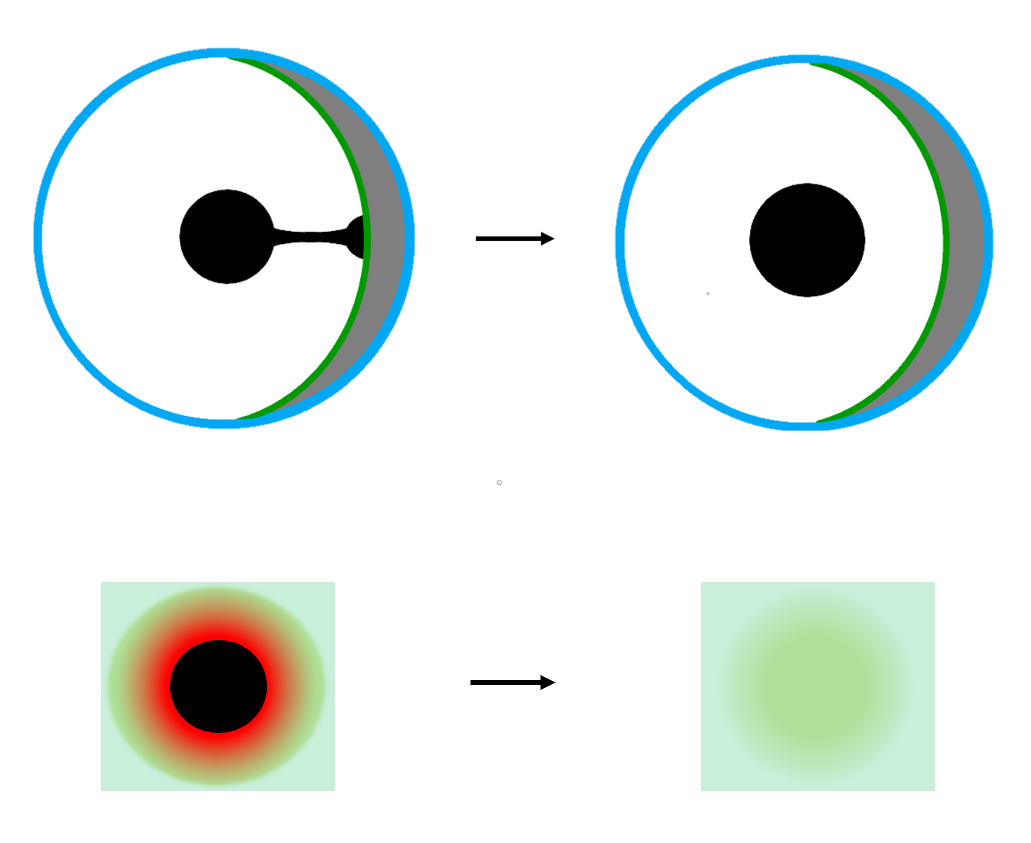}
        \caption{\small Small brane black hole connected to a large, colder black bath which induces black hole evaporation. }
        \label{fig:tumor}
\end{figure}

\begin{figure}
        \centering
         \includegraphics[width=.8\textwidth]{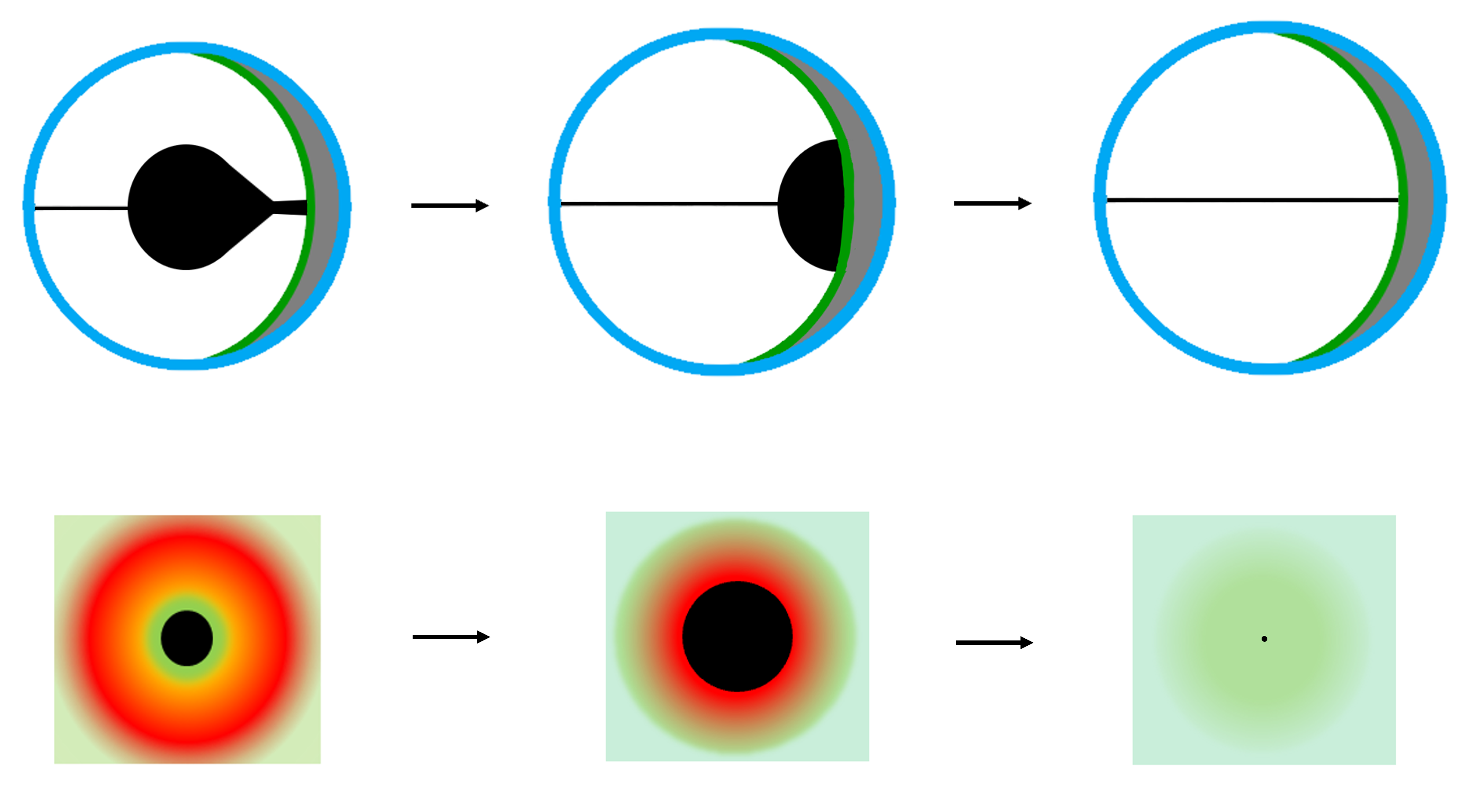}
        \caption{\small Large AdS black hole collapsing onto a black hole the brane, and subsequently evaporating into the colder asymptotic bath (an infinite reservoir), to which the system is connected by a thin but stable funnel.}
        \label{fig:three}
\end{figure}

The distinction between large and small AdS black holes will be essential for understanding the evolution where energy flows from hot to cold. One limitation of the large-$D$ effective theory is that it does not allow to combine large and small AdS black holes---that is, we can run simulations with several black holes of different sizes as long as they are all either larger than the AdS radius, or smaller than it. Within this framework, we have found find many interesting systems that exhibit holographic evaporation in each of the two classes. Out of them, we have selected the following representative examples:
\begin{itemize}
    \item Figure~\ref{fig:funneling}: A small AdS black string, suspended between two branes (in general not symmetric), is perturbed in two ways, thereby triggering: (i) the evaporation of both brane black holes into a bigger central black hole (a finite ``black tsunami''); (ii) the evaporation of one of the brane black holes onto the antipodal one, which then acts as the colder radiation bath.
    \item Figure~\ref{fig:tumor}: A small AdS black hole on the brane is connected to a bigger black hole in the bulk that acts as a colder bath. The black hole on the brane evaporates entirely into the bath black hole.
    \item Figure~\ref{fig:three}: A large, unstable AdS black droplet (formed from collapse) is connected via a thin funnel to a non-gravitating boundary cold bath. The droplet evaporates into the bath (though not completely).
\end{itemize}

Although the holographic evaporation we model is typically somewhat slower than the collapse, both processes have similar durations. This is in strong contrast with stellar-mass black holes, which collapse in a fraction of a millisecond, but their evaporation would take many Hubble times to be noticeable. The reason is that in holographic systems a huge number $\sim N^2$ of species of quanta are radiated. This dramatically accelerates the evaporation and makes it of the same parametric order in $N$ as the classical collapsing time. In dual bulk terms, both are classical processes.

Let us add that the dynamics of small AdS black holes can be quite richer than indicated above. It involves the presence of thin black funnels in the bulk, which are black-string-like horizons that can be unstable with a tendency to pinch. If the horizon pinches off to zero size in the bulk, then the connection between the brane black hole and the bath will be broken and further evaporation will be hindered. Thus, the evolution of the system depends on the competition between the rate of energy flow along the funnel---which drives the evaporation---and the rate at which the funnel pinches---which leads to a burst signal that, when it reaches the boundary, indicates the severing of the evaporation channel \cite{Emparan:2021ewh}. This is another instance of a fascinating phenomenon from the boundary perspective, which does not have any known analog in free or weakly coupled field theory. 

Finally, braneworld holography has been revisited in recent times in order to derive the Page curve \cite{Page:1993wv} followed by the entanglement entropy of the radiation emitted by a black hole to a thermal bath at the same temperature \cite{Almheiri:2019hni,Almheiri:2020cfm}. 
We will give a qualitative account of how the Page curve in our systems initially grows and later decreases, as expected for evaporating black holes, with entanglement islands appearing during the decreasing part of the curve.

In the next section, we review the large $D$ approach to these systems, introducing the effective equations, their basic solutions, and their main properties. Section~\ref{sec:collapse} shows how these methods allow describing the holographic collapse to a black hole. Section~\ref{sec:evolutions} discusses our main results, namely, the bulk evolutions for several scenarios of holographic evaporation. In Section~\ref{sec:page}, we give a qualitative discussion of how the Page curve is recovered in our setup. We then conclude with final comments in Section~\ref{sec:outlook}. Appendix~\ref{app:diagrams} is a guide to the different pictorial representations of the geometries in this article.

\section{Setup and basic properties}
\label{sec:setup}

We begin with a review of the analysis in \cite{Emparan:2021ewh,Licht:2022rke}. The starting point is the solution for a black string in AdS, which we write in Eddington-Finkelstein form as
\begin{equation}\label{adsbstring}
ds^2=\frac{L^2}{\cos^2 z}\left( dz^2 -f_{k}(r)dt^2+2 dt dr+r^2 d\Sigma^{(k)}_{n+1}\right)\,,
\end{equation}
where
\begin{equation}
f_{k}(r)=k+ r^2-\frac{r_0^n(k+r_0^2)}{r^n}\,,
\end{equation}
with 
\begin{equation}
 n=D-4   \,,
\end{equation}
and $d\Sigma^{(k)}_{n+1}$ is the metric of the $n+1$-dimensional unit sphere, unit hyperboloid, or flat space for,  respectively,
\begin{equation}
  k=\pm 1, 0\,.  
\end{equation}
The AdS radius $L$ has been scaled out of the metric, so all other quantities are in AdS units. On each section of constant $z$, including at the boundaries $z=\pm \pi/2$, we have the geometry of an AdS black hole with horizon radius $r_0$.

We will mostly be concerned with the case of $k=1$. The `planar black string', $k=0$, is recovered as a limit of the previous one when $r_0\gg 1$ and is also of interest as a slightly simpler solution. The case $k=-1$ behaves similarly but some aspects require a separate analysis which we will not perform here.

\subsection{Effective theory}

In the limit of large $D$ (i.e., large $n$), we focus on the region near the horizon and around the center of AdS, where the black string is thinner and the dynamics of interest takes place. To this effect, we change
\begin{equation}\label{limit}
\frac{r}{r_0}=e^{\rho/n}\,,\qquad z=\frac{x}{\sqrt{n}}\,,
\end{equation}
and keep $\rho$ and $x$ finite as $n\to\infty$. The resulting form of the metric \eqref{adsbstring} motivates us to look for more general solutions within the following large $D$ ansatz,
\begin{equation}\label{ansatz}
    ds^2=\frac{L^2}{\left(1-\frac{x^2}{2n}\right)^2}\left[-\left(1+\frac{k}{r_0^2}\right)A\,dt^2+\frac{2 e^{\rho/n}}{n}Udtd\rho+\frac{G^2}{n 
    }\left(dx-\frac{C}{G^2}dt\right)^2+r_0^2 e^{2\rho/n}d\Sigma^{(k)}_{n+1}\right]\,.
\end{equation}
For convenience, we have also rescaled $t\to t/r_0$. The large $n$ limit of the AdS black string can be recovered by setting
\begin{equation}\label{unifst}
    A = 1-e^{-\rho}\;,
    \qquad C=0\;,\qquad U=G=1\;.
\end{equation}
Now, allowing $A$, $C$, $U$, $G$ to be arbitrary functions of $(t, x, \rho)$ and expanding in powers of $1/n$, at leading order the dependence on $\rho$ can be solved, to yield
\begin{align}
A&=1-e^{-\rho}m(t,x)\,,& C&=e^{-\rho}p(t,x)\,,\nonumber\\
G&=1+\frac{e^{-\rho}p^2(t,x)}{2(1+k /r_0^2) m(t,x)}\frac{1}{n}\,,& U&=1-\frac{e^{-\rho}p^2(t,x)}{2(1+k /r_0^2) m(t,x)}\frac{1}{n}\;.
\end{align}
The remaining equations require that the integration functions $m$ and $p$ satisfy
\begin{subequations}\label{eq:effeqns}
\begin{align}
    \partial_t m + (\partial_x +x) (p-\partial_x m)&=0\,,\\
    \partial_t p-(\partial_x +x)\left(\partial_x p-\frac{p^2}{m}\right) + p-\left(1+\frac{k}{r_0^2}\right)\partial_x m&=0\,.
\end{align}
\end{subequations}
These are the equations of the large $D$ effective theory: by solving them we obtain a solution of the Einstein theory in AdS to leading order at large $D$. The equations are very well behaved, largely owing to their dissipative properties, and they can be easily integrated using \textsl{Mathematica}'s \texttt{NDSolve}. 

\subsection{Black strings, black holes, droplets and funnels}

The simplest solution of these equations is the black string, which, using the invariance of \eqref{eq:effeqns} under constant rescalings of $m$ and $p$, is generally obtained as 
\begin{equation}\label{unifun}
    m=m_0\,,\qquad p=0\,. 
\end{equation}
We refer to this solution as the uniform string or uniform funnel. It was shown in \cite{Emparan:2021ewh,Licht:2022rke} that when $r_0<1/\sqrt{2}$, these solutions develop a dynamical instability of Gregory-Laflamme type, which grows inhomogeneities along the string. The static zero modes at the threshold of the instability give rise to a branch of non-uniform funnels when extended beyond linearized order. They can be constructed in a perturbative expansion, but are also easy to obtain numerically for larger non-uniformity.

Another important exact solution is the ``Gaussian blob'',
\begin{equation}\label{gaussblob}
m=m_0 \exp\left(-\left(1+\frac{k}{r_0^{2}}\right)\frac{x^2}{2}\right),\qquad p=\partial_x m\,.
\end{equation}
These solutions, which are stable, can be recovered by taking the large-$D$ limit of the known AdS$_D$ black holes with horizon radius $r_0$, for all three values of $k$. This illustrates an important feature of our approach, namely, that even if our starting point were configurations of extended black strings, the effective theories also manage to capture the physics of localized black holes as solutions whose energy density vanishes at large distances, i.e., $m\to 0$ at $x\to \pm \infty$. The fact that localized black holes at large $D$ are represented by Gaussian blobs was fully exploited in \cite{Andrade:2018nsz,Andrade:2020ilm}.

Droplet solutions can be constructed numerically. They approach a black string near one of the boundaries, 
\begin{equation}
    m(x\to \infty)\to m_0\,,
\end{equation}
and then, before reaching the AdS center at $x=0$, their profile falls off approximately like the Gaussian blob. When $r_0\to \infty$, or equivalently $k=0$, we obtain a `planar black droplet', which perhaps might be simpler to obtain at finite $D$ but, as far as we know, this has not been done.\footnote{A different kind of droplets, over planar black holes in AdS$_5$, were numerically constructed in \cite{Santos:2014yja}.}

There also exist inhomogeneous droplets, which can be regarded as excited states of the CFT, and twin droplet solutions, with droplets anchored at each of the boundary ends. They can be approximated as a limit of a highly non-uniform funnel with a `ditch' in the middle.

In the following, we only discuss explicitly solutions with $k=1$. As we said, $k=0$ is recovered when $r_0\gg 1$, and many aspects of large AdS black holes are shared by $k=-1$ solutions too.

\subsection{Blobs: large and small, hot and cold}\label{subsec:blobs}

The Gaussian blob \eqref{gaussblob} illustrates how the value of $r_0$ affects the shape of the solutions: larger values of $r_0$ increase the width of the blobs. Their height is controlled by $m_0$, but, since the effective equations are invariant under constant rescalings of $m$ and $p$, the height of the solutions is meaningful only when comparing two or more of them.

Besides controlling the shapes of the blobs, $r_0$ and $m_0$ have a very relevant role in determining the temperature $T_H$ of the solutions.   This temperature was derived in \cite{Licht:2022rke}, including the first $1/n$ corrections which can be obtained from the leading order solution, yielding
\begin{equation}\label{temp}
    \frac{4\pi}{n}T_H=\frac{r_0^2+1}{r_0} + \frac1{n}\frac{r_0^2-1}{r_0^2}\ln m_0\,.
\end{equation}
This is the result for uniform black strings, but the qualitative behavior is similar for other solutions, such as the Gaussian blobs, with only minor differences in numerical factors. For our purposes here, we simply need to understand the qualitative way in which the temperature changes with $r_0$ and the height of the blob, $m_0$.

\begin{figure}
        \centering
         \includegraphics[width=.5\textwidth]{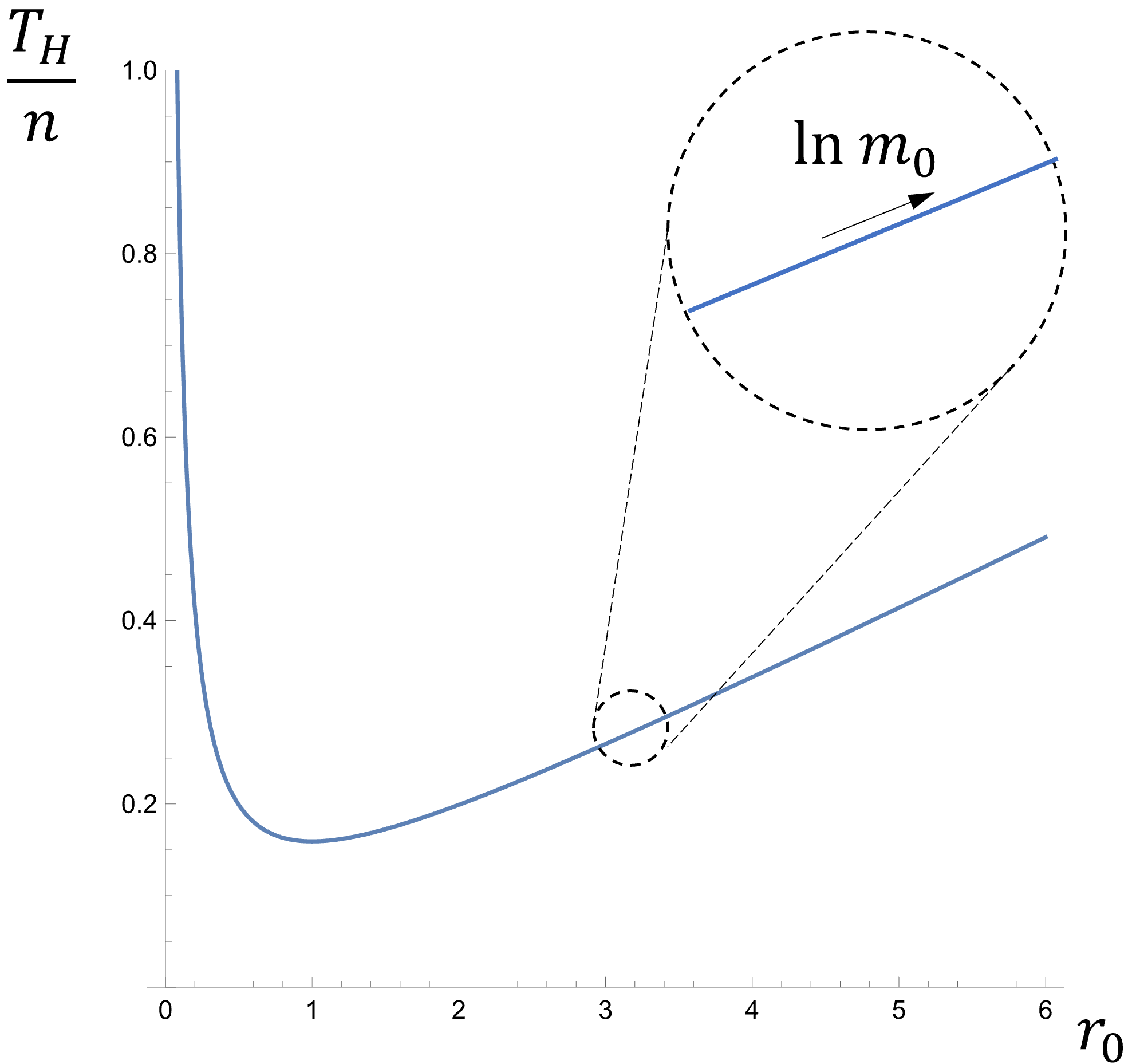}
        \caption{\small Temperature $T_H$ (normalized by $n$) as a function of blob width $r_0$ and height $m_0$ \eqref{temp}. To leading order in $1/n$, the temperature is given by $r_0$, with $T_H\sim 1/r_0$ for small AdS black holes ($r_0<1$) and $T_H\sim r_0$ for large AdS black holes ($r_0>1$). Within each of these branches, the temperature varies with the height of the blob, $m_0$, by a next order contribution $\propto 1/n$.}
        \label{fig:Tvsr0}
\end{figure}

The leading order term in \eqref{temp} gives $T_H\sim 1/r_0$  for $r_0\ll 1$ and $T_H\sim r_0$ for $r_0\gg 1$, with a minimum at $r_0=1$. This behavior reproduces the well-known distinction between small AdS black holes ($r_0<1$, with negative specific heat) and large AdS black holes ($r_0>1$, with positive specific heat). The parameter $m_0$ of the solutions adds further precision to the large/small distinction: when $r_0<1$, the temperature decreases when the size $m_0$ grows, while for $r_0>1$ it increases with larger $m_0$; see figure~\ref{fig:Tvsr0}.

In the following, we will be interested in initial configurations where we start with several black holes (i.e., blobs) with different temperatures. Since they are not in thermal equilibrium, these systems will evolve. This means that the temperature in these cases is not strictly well defined, but we can, just like in the real world, usefully associate an approximate notion of temperature to the blobs.

Since $r_0$ is not a parameter of the solutions but a parameter of the effective theory, the solutions of the equations with a given value of $r_0$ describe either large AdS black holes or small AdS black holes. Nevertheless, for a given value of $r_0$ we can combine two black holes with different temperatures, i.e, with different heights $m_0$. Since $r_0$ cannot dynamically change, its value will remain fixed in any given simulation.

With this in mind, we will follow \eqref{temp} to adopt, as a rule of thumb, that blobs with  larger height $m$
\begin{itemize}
    \item have lower temperature if $r_0<1$, 
    \item have higher temperature if $r_0>1$.
\end{itemize}
 We will then see that, in systems with several blobs, the elementary criterion that heat must flow from the hotter parts to the colder ones correctly predicts the direction of the evolution. Of course, this is nothing but the second law, and indeed the equations \eqref{eq:effeqns} governing these flows possess an entropy current with non-negative divergence \cite{Andrade:2020ilm,Emparan:2021ewh}. For our purposes here, it will suffice to discuss the effects of temperature differences.

As an example of the use of this rule, we can see that it predicts the way in which an unstable uniform string (with $r_0<1/\sqrt{2}$) evolves when perturbed\footnote{See footnote \ref{foot:GL}.}. For instance, if we add a small bulge at its center, this region will be colder and thus accrete energy from the hotter ends of the string, triggering a runaway growth of the central bulge---a black tsunami. If instead we pinch the string at the center, we create a hot `neck' which will radiate energy away from it, driving the neck thinner and eventually pinching off to zero thickness in a naked singularity \cite{Emparan:2021ewh}. If the initial uniform string has $r_0>1$, then this kind of argument concludes that these `fat' strings are stable to small perturbations. 

However, the rule must be used with some caution. For instance, static non-uniform funnels exist such that the temperature is constant on them \cite{Wiseman:2002zc,Marolf:2019wkz,Licht:2022rke}, and then it is incorrect to use \eqref{temp} to assign local temperatures to horizon regions according to their thickness. It also does not apply in the range $1/\sqrt{2}<r_0<1$, where the uniform string is stable and any small perturbation, whether creating a bulge or a neck, dissipates away. The rule, then, can break down in situations dominated by finite-size effects---that is, it strictly applies in the hydrodynamic regime of very long wavelength, but when finite scales are relevant, e.g., for non-uniform strings or strings of finite length, then it may fail. Nevertheless, when sensibly used away from these intermediate regimes, it will prove to be a useful heuristic.

\subsection{Braneworld}
Now let us assume that there is a brane at the coordinate $x=x_0$ (we take $x_0>0$ without loss of generality), so the region $x>x_0$ out to asymptotic infinity is excluded. We will work out the large-$D$ expansion of the Israel junction conditions for a brane with induced metric $\gamma_{ij}$ and extrinsic curvature $K_{ij}$, namely,
\begin{equation}\label{israel}
    K_{ij} - K \gamma_{ij}-\frac{\sqrt{D}}{L}\tau\, \gamma_{ij}=0\;.
\end{equation}
The tension $\tau$ is written here in units of $L$ and rescaled by a factor of $\sqrt D$ to retain its gravitational effect as $D\to\infty$. For a surface at $x=x_0$ in \eqref{ansatz}, expanding in $1/D$  to first non-trivial order we obtain
\begin{equation}
    K=-\frac{\sqrt{D}}{L}\,x_0\,.
\end{equation}
Along the directions of $\Sigma^{(k)}_{n+1}$ we have $K_{ij}=O(1/\sqrt{D})$, so \eqref{israel} is satisfied if we set
\begin{equation}\label{tens}
    \tau =x_0\;.
\end{equation}
Assuming this condition, and using
\begin{align}
K_{tt}&=\frac{\sqrt{D}}{2L}\frac{\partial_x m (t,x_0)-e^{-\rho}m(t,x_0) p(t,x_0)}{1-e^{-\rho}m(t,x_0)}\,\gamma_{tt}\,,
\\
    K_{t\rho}&=\frac{\sqrt{D}}{2L}\,e^{-\rho}p(t,x_0)\,\gamma_{t\rho}\,,
\end{align}
we find that \eqref{israel} requires that
\begin{equation}\label{eq:bcs}
\partial_x m (t,x_0)=0\;, \qquad p(t,x_0)=0\;.
\end{equation}

We can therefore study black holes on branes using the effective equations \eqref{eq:effeqns} and imposing the Neumann conditions \eqref{eq:bcs}.\footnote{We have also verified that the addition to the brane action of a `DGP' Einstein-Hilbert term only shifts the equilibrium tension \eqref{tens} without affecting the boundary conditions \eqref{eq:bcs}.} We will be mostly interested in situations where we have either a single brane (a Karch-Randall setup), or two branes symmetrically placed at $x=\pm x_0$ (as first considered in \cite{Emparan:1999fd}, nowadays called wedge holography). The symmetry in the latter is only for simplicity: there is no special difficulty in considering an asymmetric configuration with branes at $x=-x_L,\, +x_R$. The single-brane case can then be regarded as the limit $-x_L\to-\infty$. 

\begin{figure}
        \centering
         \includegraphics[width=1.\textwidth]{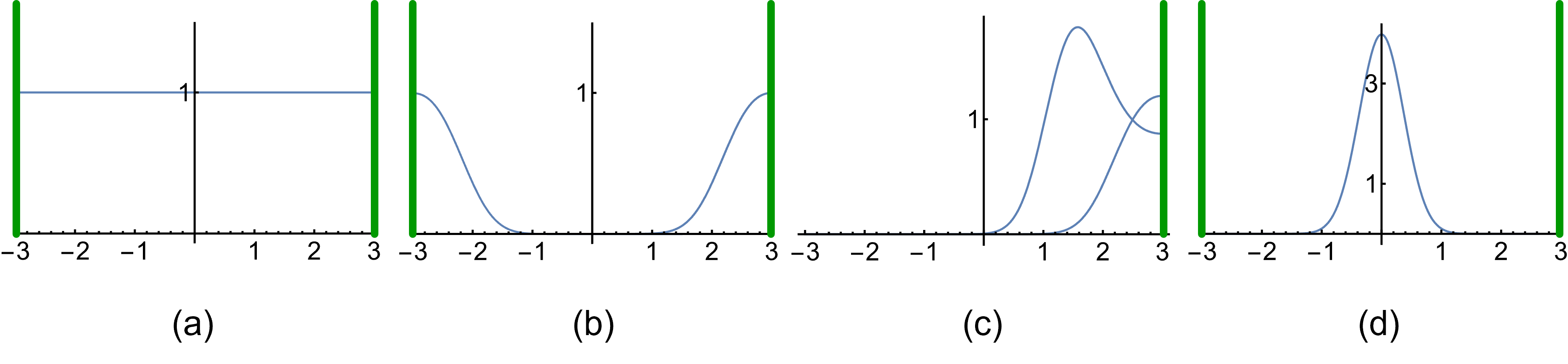}
        \caption{\small Some static solutions for black holes (blobs) in the braneworld: (a) uniform funnel; (b) double droplet; (c) single droplet, and excited-state droplet; (d) central black hole. In this and all subsequent figures, we show $m(t,x)$ and the branes are represented as green boundaries. These solutions are obtained with $r_0=.4$. Compare (a) and (c) to figure~\ref{fig:droplets and funnel}.}
        \label{fig:static}
\end{figure}

The solutions that we described above in the absence of branes are also relevant in their presence; see figure~\ref{fig:static}, and compare to figure~\ref{fig:droplets and funnel}. The simplest is the uniform funnel \eqref{unifun}, which is automatically an exact solution. Non-uniform funnels naturally appear branching out of uniform ones, with their range of existence possibly constrained by the brane boundaries. Indeed, the presence of branes that limit the extent of the black string can enhance their stability. The stability bound $r_0>/1/\sqrt{2}$ applies for an infinite string, but a string with $r_0<1/\sqrt{2}$ will be stable when suspended between two branes closer than the wavelength of the lowest unstable mode (so the rule in Sec.~\ref{subsec:blobs} is invalidated due to finite-size effects).

Gaussian blobs centered at $x=0$ are not exact solutions now since their profile will be modified near the brane to satisfy the boundary conditions \eqref{eq:bcs}. They will have non-zero amplitude on the brane, $m(x_0)$, and if this amplitude is exponentially small, then the blob will be very close to the Gaussian and may be regarded as separate from the brane. This is likely the correct interpretation when $r_0<1/\sqrt{2}$, since in this case, thin horizons are unstable to pinching. If the amplitude on the brane is larger, then this must be interpreted as the blob giving rise to a horizon on the brane. There is no clear-cut distinction between the two cases, and sometimes this ambiguity will affect the choice of interpretation.

The amount of distortion of the Gaussian shape depends on the relative values of the width of the blob, $r_0$, and the distance to the brane, $x_0$.
This feature is even more relevant for droplets, which are localized away from the center. When $x_0$ is large and/or $r_0$ is small, one readily finds static stable droplets localized on the brane. But these will stop existing when their width $r_0$ is comparable or larger than $x_0$, since in that case, the brane does not leave enough room to `fit' them to one side of the AdS center.

\subsection{Braneworld gravity at large $D$}\label{subsec:bwlargeD}

It is well known that dynamical gravity is induced on a braneworld model in AdS$_{D}$ \cite{Randall:1999vf,Karch:2000ct}. When the brane is close to the asymptotic AdS boundary, the effective theory on the brane is well described by $D-1$-dimensional Einstein gravity (with a graviton mass if the brane has AdS$_{D-1}$ geometry).

In our construction, with \eqref{limit}, we focus on a region near the AdS center, since the dynamics of the horizon is localized there. This means that the effective theory on the brane receives large corrections, as can be confirmed by an explicit analysis of the higher-order terms. Hence, it does not reproduce well gravity in $D-1$ dimensions and, instead, gravity on the brane is closer to a $D$-dimensional theory. This is a limitation of the large $D$ approach to this problem. Nevertheless, as we will see, although our results may not be quantitatively accurate, the conclusions appear to be physically sensible and qualitatively sound. 

With this framework in place, in the next two sections we present the results of the numerical solutions of the effective equations for several phenomena. These calculations provide the basis for the pictorial accounts that we gave in the introduction.

\section{Collapse on the brane}
\label{sec:collapse}

\begin{figure}
        \centering
         \includegraphics[width=1.\textwidth]{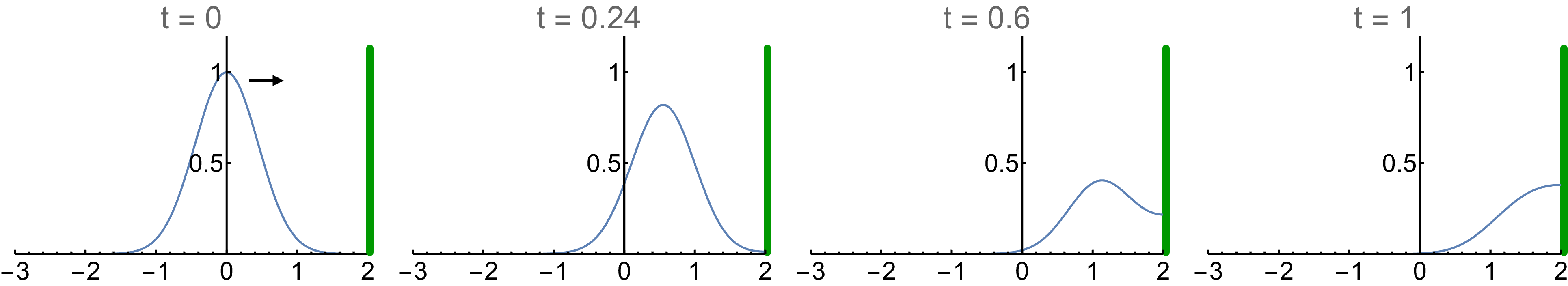}
        \caption{\small Evolution of collapse to form a black hole on the brane. An initial bulk black hole, shown here as a Gaussian blob, is given a boost towards the brane. When the blob hits the brane, it sticks to it, forming a stable droplet. This simulation is performed with $r_0=.5$ and initial velocity $3$. Compare to figure~\ref{fig:collapsing}.}
        \label{fig:collapse}
\end{figure}

The first process that we describe is physically sensible and clear  (cf.~Figs.~\ref{fig:collapsing} and \ref{fig:collapse}). We take a black hole in the bulk, away from the brane, and give it a boost towards the brane (the boost transformation for blobs was presented in \cite{Licht:2022rke}). The bulk black hole hits the brane and sticks to it, thus forming a black droplet, which is a stable configuration since it does not have a funnel through which it would evaporate. In dual terms, we start with a spherical cloud of thermal plasma and give it a radial inwards push. The cloud then collapses to form a black hole, which is surrounded by a stable halo of quantum conformal fields.

This collapse is most easily attained in the regime $r_0<1$ of small AdS black holes, since these readily form stable droplets. In Sec.~\ref{subsec:collevap} we present another instance of collapse involving large AdS black holes, with $r_0>1$. In that case, we will see that the black hole does not remain static but proceeds to evaporate.


\section{Evaporation on the brane}
\label{sec:evolutions}

Now we study configurations that start with a black hole on the brane. We want to see this black hole evaporate. As we discussed, the evaporation corresponds to the horizon shrinking on the brane and sliding off into the bulk, in such a way that the bulk horizon area does not decrease. The latter is guaranteed by the bulk equations.

Static black droplets do not serve this purpose, even if we give them a small kick: they are stable configurations. To enable evaporation, there must exist a channel for the horizon to flow away from the brane into the bulk. In the following, we discuss different implementations of this idea. The first three scenarios correspond to small AdS black holes, with $r_0<1$, and the last one to large AdS black holes with $r_0>1$. 

\subsection{Twin black hole evaporation}\label{sec:twin}

\begin{figure}
        \centering
         \includegraphics[width=.9\textwidth]{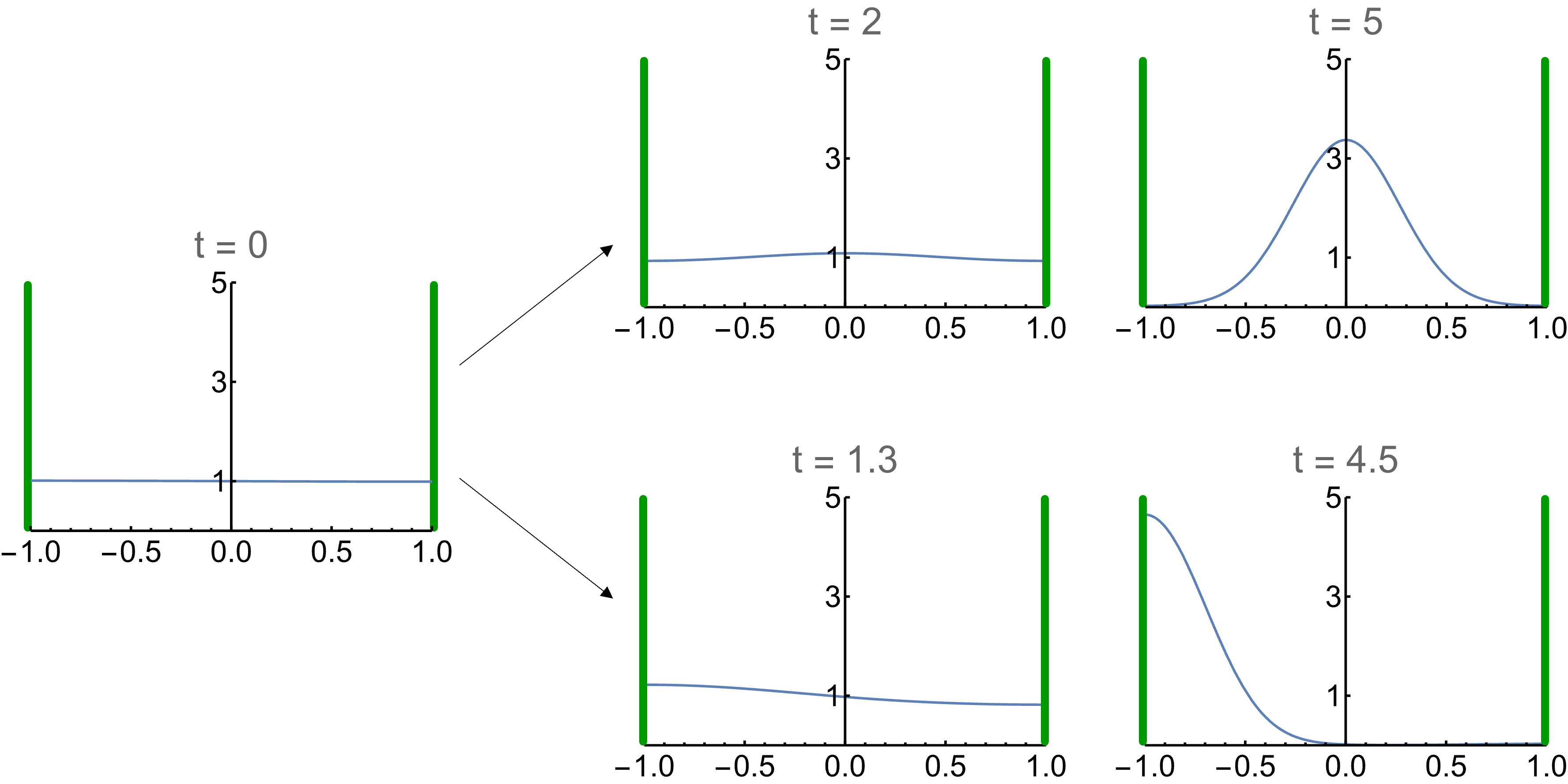}
        \caption{\small Alternative evolution paths for the unstable uniform funnel. Up: when the perturbation forms a central bulge, a black tsunami flows from the brane to the bulk, ending at a central black hole without any horizons on the branes. Down: an asymmetric perturbation makes the horizon flow away from one of the branes towards the other. In both cases, the direction of evolution can be inferred from the fact that, since these are small AdS black holes, the thinner parts of the horizon are hotter and radiate towards the thicker, colder regions. Both simulations are made with $r_0=0.27$. Compare to figure~\ref{fig:funneling}.}
        \label{fig:evaporationtwin}
\end{figure}

Consider a uniform black string suspended between two symmetrical branes (cf.~Figs.~\ref{fig:funneling} and \ref{fig:evaporationtwin}). We take a thin string with $r_0$ sufficiently less than $1/\sqrt{2}$, so it is unstable. We then perturb it by putting a bulge around its center at $x=0$. The evolution is as discussed at the end of Sec.~\ref{subsec:blobs}: a Gregory-Laflamme instability is triggered, which makes the central bulge grow. This growth is similar to the black tsunami flow in \cite{Emparan:2021ewh}, but, since the total energy of the system is conserved\footnote{This can be derived from the effective equations \eqref{eq:effeqns} with the Neumann conditions \eqref{eq:bcs}.}, the string must become thinner away from the center, and therefore the horizons on the branes shrink. Eventually, we end with a Gaussian bulge at the center of AdS, and no horizons on the branes. This evolution is shown in the upper part of figure~\ref{fig:evaporationtwin}.

Here we recognize at work the general principle of heat flow. Since we are in the regime of small AdS black holes, the initial bulge is colder than the string endpoints, and thus heat flows, in a runaway, towards the center. In dual boundary terms, we start with two black holes at the antipodes of a spatially spherical universe. The holographic CFT is in the peculiar state dual to a uniform funnel: its stress tensor is zero but there is non-trivial $\ord{N^2}$ entanglement between the two black holes. Since $r_0$ is small, this is an unstable state of the CFT. We introduce a perturbation that cools down the CFT in the region away from the black holes, and also makes these a little smaller and thus hotter than their surroundings. The black holes then evaporate away, leaving a thermal state of the CFT filling the universe. 

The latter is the end state in the limit $N\to \infty$; if finite $N$ corrections are included, then the central black hole would evaporate through bulk quantum radiation. We have also assumed that the branes are well separated, i.e., $x_0$ quite larger than $r_0$.  If instead, $x_0$ is smaller than $r_0$, the instability may be quenched and the black holes on the branes would remain. We shall not attempt here to demarcate in any detail the regions of initial parameters that lead to specific final outcomes.

\subsection{One black hole radiating into another}

We can obtain another evolution of the same state of an unstable black string suspended between two branes if we apply a different initial perturbation. This time, we will make the string thinner at one endpoint and thicker at the other. The Gregory-Laflamme instability is now triggered in such a way that the thinner end shrinks and the thicker one grows, as shown in the lower part of figure~\ref{fig:evaporationtwin}. Again, the evolution conforms to the notion that heat flows from the hotter (thinner) part of the bulk horizon to the colder (thicker) part. 

In the dual boundary view, the two antipodal black holes start with slightly different temperatures. The CFT is in a state that allows the transfer of heat between them involving $\ord{N^2}$ degrees of freedom. As a consequence, the hotter black hole quickly radiates all of its energy into the colder one.

Similar considerations apply as before concerning the effects of finite $N$ and small inter-brane distances.

\subsection{Black hole evaporating into a colder bath}\label{subsec:coldbath}

\begin{figure}
        \centering
         \includegraphics[width=1.\textwidth]{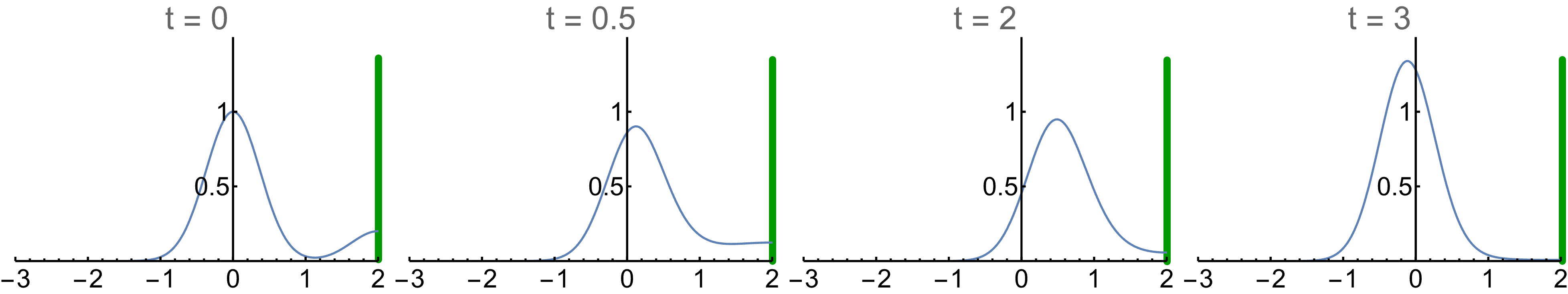}
        \caption{\small A black droplet on the brane is connected to a larger and colder black hole in the bulk. The initial system is far from equilibrium and the droplet is quickly absorbed by the central black hole (which then wobbles around the AdS center). Here we have set $r_0=0.4$. Compare to figure~\ref{fig:tumor}.}
        \label{fig:evaporationbath}
\end{figure}

The previous cases started with configurations that were slight perturbations of unstable equilibrium states. In the next example (cf.~Figs.~\ref{fig:tumor} and \ref{fig:evaporationbath}), we are still in the regime of small AdS black holes, but the initial states are not close to equilibrium. Instead, we set up a single-brane configuration with two blobs, a smaller one localized near the brane, and a bigger one in the bulk, with a funnel joining them. For the initial blobs, we can simply take the solutions for a droplet on the brane and a Gaussian centered at $x=0$; if the tails of their profiles reach each other appreciably, this is enough for a funnel-like connection between them. 

Even though separately each initial blob may be a static solution of the equations, when combined, they will not remain static. We expect that the smaller (hotter) blob on the brane flows to the larger (colder) blob in the bulk, and this is indeed what the numerical evolution of the equations shows, see figure~\ref{fig:evaporationbath}.

In dual terms, the black hole on the brane evaporates into a colder bath. This bath is a large non-gravitational system, namely the CFT at the asymptotic AdS boundary\footnote{The evolution would be essentially the same if there were a second brane at large negative $x$.}, and the bulk black hole sets it at a finite temperature that we take to be lower than the brane black hole. The overlap between bulk blobs provides the channel for the classical flow of horizon generators (dual to heat) between them.

\subsection{Quick collapse followed by slow evaporation into an empty bath }\label{subsec:collevap}

\begin{figure}
        \centering
         \includegraphics[width=1.\textwidth]{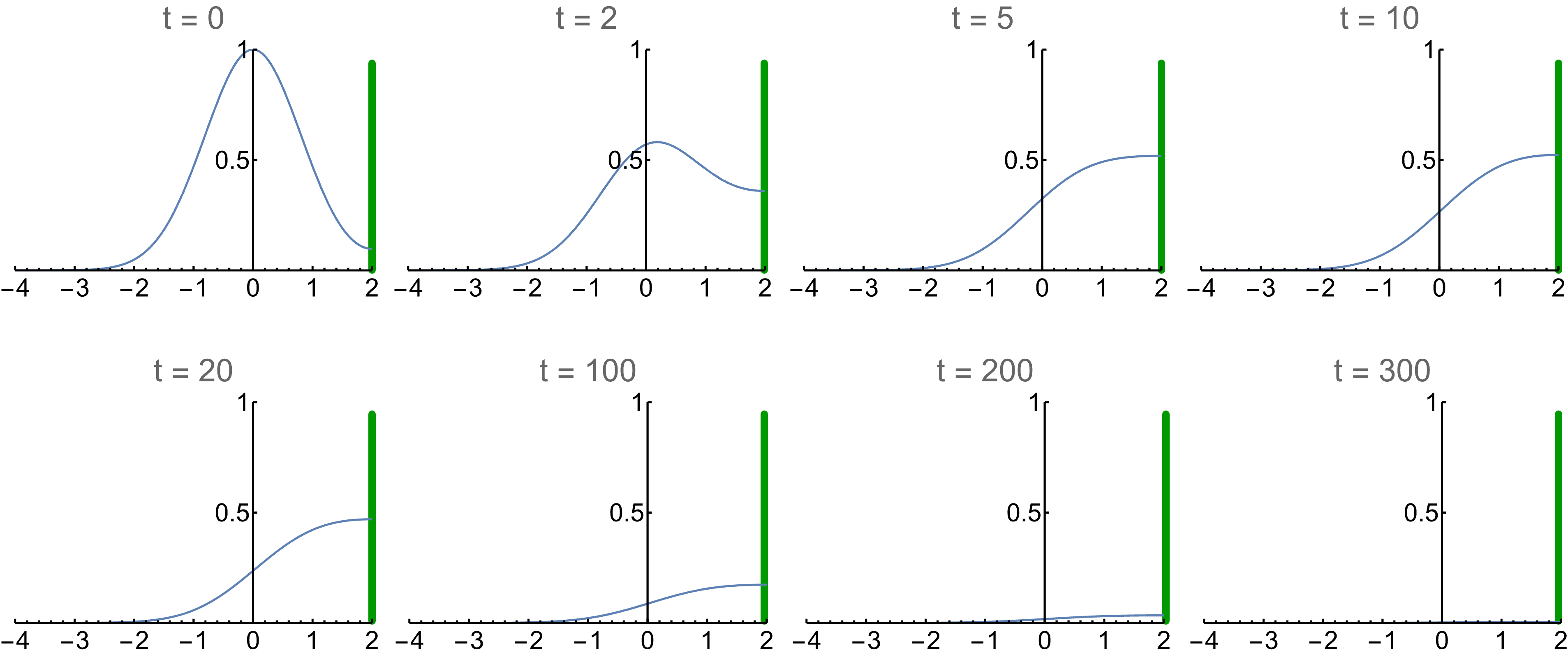}
        \caption{\small Upper row: Quick collapse of a central large AdS black hole onto a smaller one on the brane, forming an unstable black droplet. Lower row: The droplet slowly slides towards the bulk. At all times the horizons are connected through a thin (but stable) funnel to the leftmost asymptotic boundary, so that, on this non-dynamical boundary, there is a colder black hole that can absorb (arbitrarily) large amounts of heat. Observe that while the collapse happens in a few units of time, the evaporation is slower. The reason is that the temperature, and hence the radiation rate, of large AdS black holes decreases as their size becomes smaller. Here $r_0=1.4$. Compare to figure~\ref{fig:three}.}
        \label{fig:evaporationlarge}
\end{figure}

Now we consider large AdS black holes, i.e., solutions of the effective equations with $r_0>1$ which are thermodynamically stable. The (approximately) Gaussian profiles of the corresponding blobs are now broader. We have not found conclusive evidence of the existence of static droplets in this range, at least not for branes with moderate values of $x_0$. Regardless of this, we can always set up initial conditions for a blob localized on the brane, which does not settle into a static droplet and which then evolves by sliding off into the bulk. That is, we set up initial conditions for a non-equilibrium configuration of a black hole on the brane surrounded by a CFT halo. This black hole will subsequently evaporate (cf.~Figs.~\ref{fig:three} and \ref{fig:evaporationlarge}).

In our simulations, we have modeled the entire process of collapse and evaporation. We start from a large black hole in the bulk that has a substantial overlap with the brane. This is not a stable configuration; the bulk black hole moves towards the brane, and for a few units of time, it sticks there; see the upper row of figure~\ref{fig:evaporationlarge}. In dual terms, we start with a large, hot cloud of CFT plasma that surrounds a small, colder black hole. The cloud collapses onto the black hole, yielding a larger, hotter black hole, surrounded by a large CFT halo.

Then, dual evaporation occurs: the unstable droplet begins to be sucked into the bulk. Where does its energy go into? At this point, it is important to realize that we are in the regime $r_0>1$ where uniform funnels are stable. At all times, our system has a tail that extends out to the asymptotic boundary at $x\to -\infty$. We argued that when $r_0<1$ it seems adequate to think that the exponentially small tail does not represent a connection of the bulk horizon to asymptotic infinity: this would be a thin unstable string that would pinch off. But when $r_0>1$ such a string would be a stable funnel that enables the transfer of heat to the colder bath at the non-gravitating boundary. That is, our system is always connected to a (non-dynamical) bath with a black hole in it. The droplet on the brane then emits all its energy into this colder black hole, as shown in the lower row of figure~\ref{fig:evaporationlarge}.

\subsection{Remarks}\label{subsec:remarks}

Holography classicalizes the process of Hawking evaporation, and this introduces several peculiarities as compared to the situation where only a few, weakly coupled quantum fields are radiated.

\paragraph{Evaporation rate.} One feature of the holographic setup is that, as we have seen, the collapse and the evaporation can be modeled in a single simulation without a large disparity in their time scales. Since both are classical bulk processes, they are of the same parametric order in $N^2$. Usually, the quantum evaporation of a black hole is much slower than the collapse, but here the evaporation rate is hugely enhanced by the large number of degrees of freedom of the CFT that it radiates. Still, in the simulations with large AdS black holes in Sec.~\ref{subsec:collevap},  one observes that the collapse happens more quickly than the evaporation, which indeed slows down in its late stages. This is a very minor effect relative to the large $N$ enhancement, and it is due to the fact that these are large AdS black holes. From the viewpoint of the brane, they evaporate because they are coupled to a (non-gravitating) bath. But their temperature is lower the smaller they are, so, in contrast to small AdS black holes, their radiation rate slows down as they evaporate.

\paragraph{Endpoint of evaporation.} Holographic evaporation follows a classical process of horizon shrinking and pinching that belongs in the class of Gregory-Laflamme horizon instabilities. Thus, the outcome of the holographic evaporation will be dual to the fate of the Gregory-Laflamme instability of black strings. There is strong evidence that, in any finite number of dimensions, thin enough black strings pinch off in a finite time \cite{Lehner:2010pn,Emparan:2018bmi}, and this is what we expect to happen for $r_0\lesssim 1/\sqrt{2}$ and $x_0$ not too small (recall that small $x_0$ can enhance stability). The simulations that we have presented above are such that the horizon pinches off on the brane. In this case, the endpoint of Hawking evaporation is described by the same physics as the endpoint of the Gregory-Laflamme instability. 

\begin{figure}
        \centering
         \includegraphics[width=1.\textwidth]{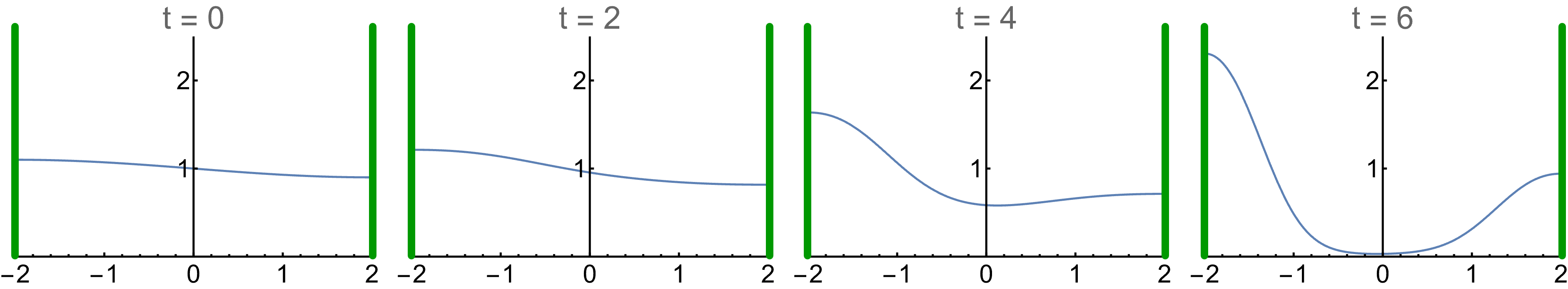}
        \caption{\small Evolution of a perturbed uniform funnel where a singular pinch occurs in the bulk faster than the heat transfer across the funnel. This can be seen as the competition between two unstable Gregory-Laflamme modes of the black string, such that it pinches in the bulk instead of on the brane. In the dual view, the entanglement between the two black holes is severed, yielding a burst of radiation that stops the evaporation. In this simulation, $r_0=0.4$.}
        \label{fig:pinch}
\end{figure}

However, it is also possible to devise that the horizon pinches off not on the brane but in the bulk. We can achieve this by choosing different initial perturbations of an unstable uniform funnel, but also starting with different sets of blobs with $r_0\lesssim 1/\sqrt{2}$. When the `neck' connecting the two blobs is long and/or thin enough there is a competition between the rate at which heat flows across the neck and the rate at which the neck shrinks. This can be achieved by exciting different unstable Gregory-Laflamme modes of the black string. In the event that the neck pinches off to zero before the evaporation is complete, the channel that entangles the black hole to the radiation bath is severed and the evaporation stops.  The endpoint is a droplet on the brane and a black bath, or possibly a droplet at another brane, both with different temperatures. We illustrate an instance of this in figure~\ref{fig:pinch}. The sudden burst of CFT radiation that precedes the end of the evaporation was studied in \cite{Emparan:2021ewh}. This is an extremely peculiar phenomenon from the boundary viewpoint.

\paragraph{Thermal equilibration.} Note also that the horizon need not always pinch-off to zero, either in the bulk or on the brane. That outcome can be averted by either having large enough black holes or funnels, $r_0\gtrsim 1/\sqrt{2}$, or having a small enough separation in a double-brane wedge system. The endpoint of these evolutions will then be a funnel, a uniform one or possibly (but more rarely) a non-uniform one. In these cases, we start with a black hole that is connected to a colder bath, and the evolution ends when the black hole reaches thermal equilibrium with the bath, in a Hartle-Hawking-like state or in the peculiar state dual to the uniform funnel. It is also possible to have, and easy to model with our methods, the converse process of a black hole absorbing radiation from a hotter bath until both systems come into thermal equilibrium.

\paragraph{Finite $N$ evaporation.} Finally, it should be clear that the scenarios that end with a stable droplet are not true endpoints when $N$ is large but finite, but only long-lived phases. Emission of color singlet states (e.g., glueballs) is possible when $N<\infty$, and evaporation will happen through Hawking radiation of bulk quanta in a time that scales as a power of~$N$.


\section{Page curve for evaporating black holes}
\label{sec:page}

Work in recent years has given us a better understanding of the entanglement structure between a black hole and its radiation \cite{Penington:2019npb, Almheiri:2019psf, Almheiri:2020cfm}. 
Further support 
has been obtained through braneworld models of black hole evaporation, 
which allow using the classical dual for entropy calculations \cite{Almheiri:2019hni}.

\paragraph{Previous setups.} Let us briefly review the braneworld island picture. The black hole is placed on the brane, and the boundary CFT plays the role of the bath. 
This point of view, where the brane black hole is put in the spotlight, is known as the \textit{intermediate picture} setup, as opposed to the complete UV picture (the BCFT) and the pure gravity picture (the higher-dimensional bulk). Given the nature of this doubly-holographic setup, the brane black hole is immersed in a strongly-coupled, large $N$ CFT which backreacts on the brane geometry. When the brane black hole is eternal, the Page curve features an initial growth in the entropy, followed by saturation at $2S_{BH}$, where $S_{BH}$ is the Bekenstein-Hawking entropy. The change in the slope is realized through the quantum extremal prescription,
\begin{equation}\label{qes}
    S(R) = \text{min}\left\{\underset{I}{\text{ext}} \left[ \frac{\text{Area}(\partial I)}{4 G_N \hbar} + S_{\text{bulk}}(I\cup R)\right] \right\},
\end{equation}
where $I$ is an island, $\partial I$ is its boundary, $R$ is the radiation, and $S_{\text{bulk}}$ is the entanglement entropy of the island together with the radiation, computed semiclassically. Entanglement entropies are difficult to compute directly, especially in dimensions higher than two, which is why we resort to the bulk computation where (H)RT surfaces geometrize the answer. From the brane perspective, the island phase emerges once the minimal RT surfaces in the bulk extend across the brane.
This geometric connection is related to the entanglement between the Hawking radiation and the interior of the brane black hole.

The pre-Page time system increases in entropy, due to the growing size of the ER bridge 
\cite{Hartman:2013qma, Neuenfeld:2021bsb}. However, when the exterior of the black hole is static, the island RT surface gives a time-independent contribution which eventually dominates $S(R)$ and explains the saturation of the Page curve. Here we have in mind the construction of \cite{Almheiri:2019yqk}, where the radiation system and the black hole are eternally connected. In this case, the island extends outside the black hole horizon.\footnote{If a mechanism is introduced to connect the bath and the black hole only at some finite time, the island appears inside the horizon \cite{Almheiri:2019hni}. While it might be possible to implement the latter in our setup, we will not attempt it in this article.}

\paragraph{Our setup.} How is our setup different from previous constructions in $D\geq 5$? The main difference is that those setups involved static black holes in a situation of thermal equilibrium---what changes in them is the structure of entanglement, but not the stress tensor of the radiation nor the geometry of the black hole. 
Instead, our systems start away from thermal equilibrium and dynamically evolve in time, 
with evaporation that can be followed until almost the very end. 

However, one aspect in which our setups are \emph{not} different from previous ones is that our bulk black holes are not formed from collapse nor disappear into bulk radiation, but are, in that sense, eternal. That is, collapse and evaporation only occur from the point of view of the brane. An eternal bulk black hole describes the dual CFT radiation as a thermofield double, connecting two separate asymptotic boundary regions, and the state of the CFT in a single boundary component is not pure. This is also the case in our configurations, which can be regarded as dynamical versions of a thermofield double. Then we expect that, as in \cite{Almheiri:2019yqk}, the islands reach outside the horizon.

Other differences concern the type of black holes in the higher-dimensional bulk. In \cite{Chen:2020uac,Chen:2020hmv} they are AdS-Rindler spacetimes, which can be regarded as the simplest uniform funnels, and are the solutions \eqref{adsbstring} with $r_0=0$ and $k=-1$, which are not amenable to our large $D$ study. Other solutions, built numerically \cite{Almheiri:2019psy}, describe equilibrium funnels in a Poincar\'e patch Randall-Sundrum braneworld in AdS$_5$. Clearly, any equilibrium configuration where there is entanglement between the black hole and the radiation must have a static or stationary funnel.\footnote{The model in \cite{Geng:2020qvw} can also be regarded as connecting the black hole and the bath in equilibrium.}

Our setups introduce dynamical evolution in the picture. Due to the time dependence, the relevant entropies must be computed through the quantum HRT formula. In principle, this can be evaluated for our solutions since we have the complete form of the metrics, even if in numerical form. However, the actual computations are delicate and outside the scope of this article. Here we will remain content with qualitatively arguing for the dominant surfaces through the holographic form of the quantum extremal prescription.

Let us discuss the Page curve for the evaporation of a droplet into a colder bath, which we examined in Secs.~\ref{subsec:coldbath} and \ref{subsec:collevap}. For this purpose, it will be convenient to resort to a different kind of diagram that represents the two asymptotic regions of the geometry (see Appendix~\ref{app:diagrams}). For instance, figure~\ref{fig:rattle} describes the same configuration as figure~\ref{fig:tumor}, but the latter only showed one of the two asymptotic regions.
\begin{figure}
        \centering
         \includegraphics[width=.6\textwidth]{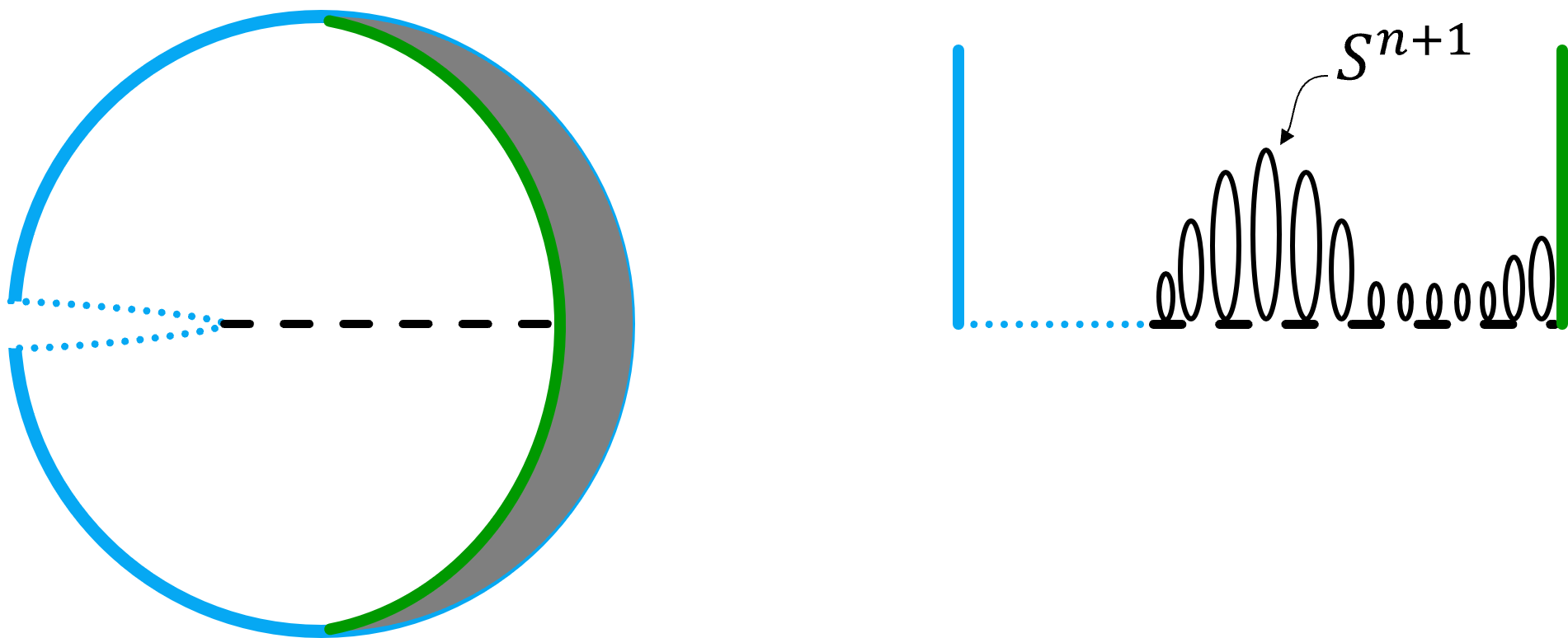}
        \caption{\small A different representation of the initial configuration in figure~\ref{fig:tumor} (cf.\ App.~\ref{app:diagrams}). \textbf{Left}: Each point in this Cauchy slice is a sphere $S^{n+1}$. The Einstein-Rosen bridge that connects the two asymptotic regions is represented as a black dashed line. The blue dotted line is the $SO(n+2)$ rotation symmetry axis (in each region). \textbf{Right}: the $S^{n+1}$ remain of finite size at the ER bridge, but shrink to zero on the symmetry axis. The radius of these circles is represented in figure~\ref{fig:tumor}, which only shows one asymptotic region.}
        \label{fig:rattle}
\end{figure}

\begin{figure}[t!]
        \centering
         \includegraphics[width=.9\textwidth]{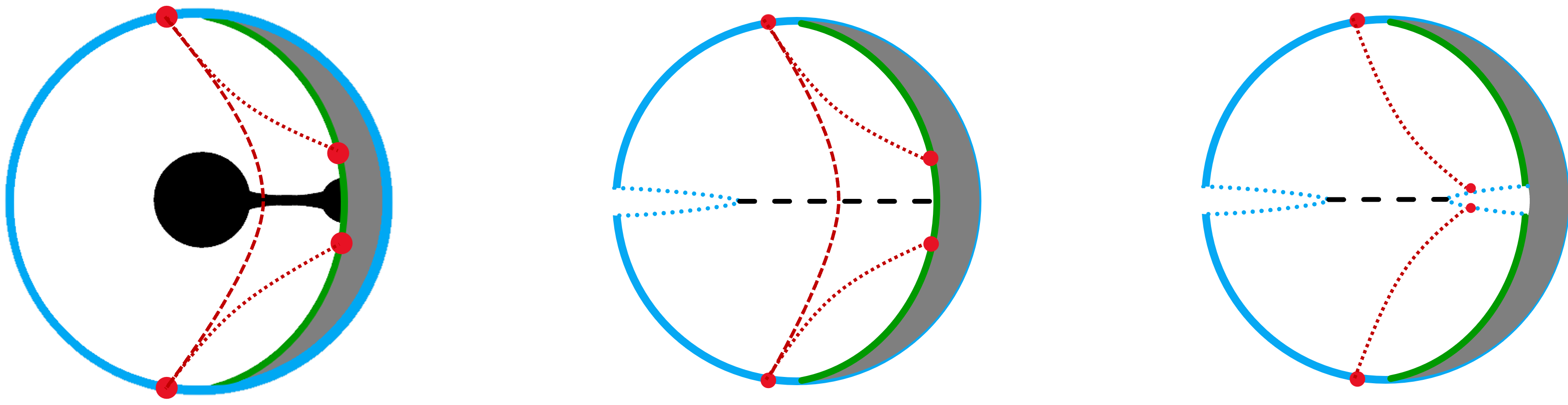}
        \caption{\small HRT surfaces for the calculation of the Page curve in an evaporating droplet: \textbf{Left}: surfaces (red dotted and dashed)  in the representation of figure~\ref{fig:tumor}, which should be thought of as rotated along the horizontal central axis, and only shows the half of the surfaces in one of the asymptotic regions. \textbf{Middle}: surfaces in the representation of figure~\ref{fig:rattle}, which shows the two asymptotic regions. \textbf{Right}: the situation after the black hole has evaporated from the brane. We are omitting HRT surfaces that would pass to the left of the black hole, which could be dominant if the bath saturates its entanglement.}
        \label{fig:rts}
\end{figure}

With this understanding, in figure~\ref{fig:rts} we draw the HRT surfaces on a given Cauchy slice, in both kinds of diagrams.
The HRT surface is anchored in the non-gravitating bath, and in the early moments, it should go directly through the horizon. 
Without a detailed calculation, we cannot pinpoint the precise place in the bulk horizon where the HRT surface enters, but we may presume (assuming, e.g., time-symmetric initial data) that initially, this will happen near where the horizon is thinner and the ER bridge is smaller. This initial quantum extremal surface will move as the bulk horizon evolves, but its increase will come mostly from the growth of the ER bridge. This gives the initial rise in the Page curve for the entropy.

At some moment, this growth will lead to a change in dominance of the quantum extremal surface. Namely, the area of the HRT surface will be minimized by an island surface that ends on the brane, near the horizon of the brane black hole. However, since our brane black hole shrinks, this island will not give a constant contribution to the entropy, but a decreasing one, following the horizon of the brane black hole. Thus, the decrease in the Page curve results from the shrinking of the droplet. When, eventually, the brane black hole completely evaporates, it leaves the initial value for the entropy (up to a small, $1/D$ increase due to total bulk horizon growth), which we can subtract away.

Figure.~\ref{fig:rts} (right) presents the bulk geometry when the evaporation on the brane is complete. For this to occur, the bulk horizon must split through a topology-changing process that involves the resolution of a mild naked singularity.

\paragraph{Finite bath effects.} In this setup, a third kind of HRT surface is also possible, which passes to the left of the horizon without intersecting it. Like the island surfaces, these ones vary in time only because the whole geometry is time-dependent, but this variation can be expected to be relatively small. Surfaces of this type appear because the bath is finite (the bulk black hole has finite size) and they were identified as such in \cite{Grimaldi:2022suv}. If these surfaces become dominant at some point, they signal the saturation of the bath in a thermalized phase. The Page curve that we described above, which grows and then decreases to zero, assumes that the bath (i.e., the bulk black hole) is large enough that this saturation does not happen.

Observe that a static black droplet has the same topology as the black hole in figure~\ref{fig:rts} and can be regarded as a special case of it. Indeed, a small droplet will slightly `heat up' the non-gravitating boundary region. The degrees of freedom excited in this region are in a `partially deconfined phase' of the CFT of the type that is dual to a small AdS black hole.

\begin{figure}
        \centering
         \includegraphics[width=.5\textwidth]{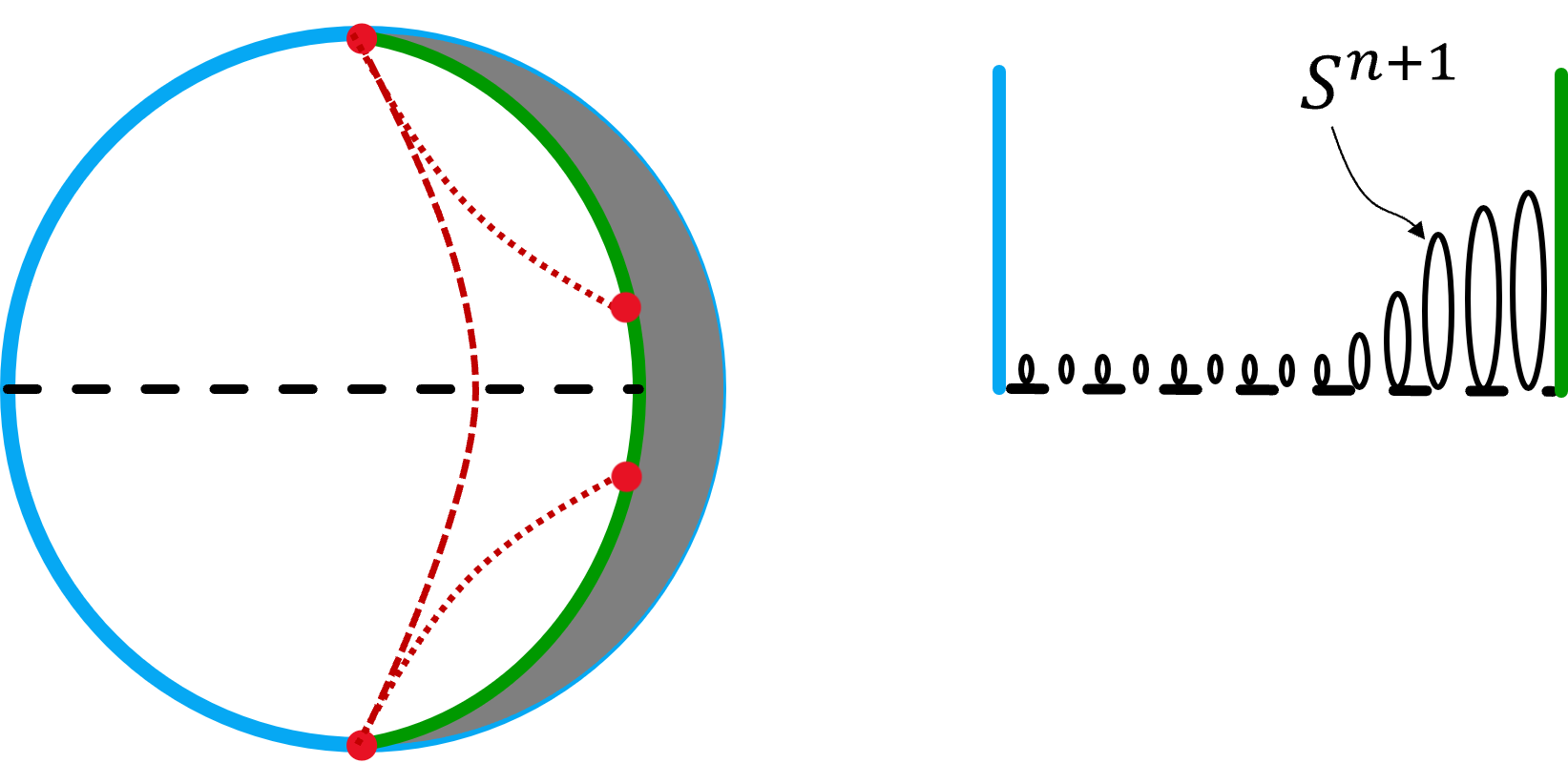}
        \caption{\small A different representation (analogous to figure~\ref{fig:rattle}) of the configuration in figure~\ref{fig:three} (middle) for evaporation into an infinite bath. We also show the HRT surfaces in it. This shares qualitative features with the scenario in \cite{Almheiri:2019hni}, in that the Page curve at late times plateaus at a constant value, but at earlier times its behavior is more complex, first growing and then decreasing until reaching the plateau.}
        \label{fig:infbath}
\end{figure}

\paragraph{Infinite baths.} Figure~\ref{fig:infbath} depicts the geometries for scenarios with an infinite bath, such as we considered in Sec.~\ref{subsec:collevap}. In this case, the Page curve drops during the evaporation but, since the evolution stops at a stable uniform funnel, eventually the curve plateaus at a constant value, as in the scenarios in \cite{Penington:2019npb, Almheiri:2019psf, Almheiri:2020cfm, Almheiri:2019hni}.

\paragraph{Time scales.} Let us examine how the large $D$ limit affects the time scales involved\footnote{These differ markedly from the scales for evaporation into a few weakly coupled fields at large $D$ \cite{Holdt-Sorensen:2019tne}.}. As we discussed above, in holographic braneworld setups the characteristic evaporation time, and hence the Page time, are of the order of the classical bulk AdS scale. This is still the case in our models even if $D\to\infty$. Note, however, that in \cite{Chen:2020hmv} the Page time was found to obey
\begin{equation}\label{paged}
    \tau_p \sim \frac{(D-1)^{\frac{D-1}{2}}}{(D-2)^{D/2}} \frac{1}{\theta^{D-2}}\,,
\end{equation}
where $\theta$ is the angle of the brane, which is inverse in the brane tension. This relation was derived for small angles, so for cases where the brane is very close to the boundary. Clearly, as we take the brane to the boundary, the Page time becomes larger (owing to the fact that the Planck length on the brane becomes smaller), and when $\theta<1$ it would seem to diverge as $D\to\infty$. Eq.~\ref{paged} was derived for the eternal black hole setup and it is not immediately obvious that it applies to dynamically evolving systems. However, what makes our scenarios most different is that, as we discussed in Sec.~\ref{subsec:bwlargeD}, our branes are not close to the boundary, so $\theta$ is not small but rather of order one. 

When $D$ is large, the Gregory-Laflamme-type dynamics that drives the evolution of the bulk horizon, with a characteristic time of the order of the AdS length (for $r_0$ parametrically of that order too), is concentrated near the AdS center. By placing the brane in this region, the evaporation can be completed in a few AdS time units, as we have observed. At finite $D$ this can also occur when the brane is further away, since the dynamical evolution is expected to involve an exponentially growing black tsunami flow \cite{Emparan:2021ewh}, which can quickly drive the evaporation within an AdS time scale. 

\paragraph{Truncated evaporation.} Other interesting evolutions are revealed within our framework which are expected to occur also when $D$ is finite. We already discussed that, when the droplet is connected to the bulk horizon via a thin, unstable black string, the Gregory-Laflamme instability of the latter can win over the evaporation. The pinching of the string prevents the brane black hole from accessing the necessary degrees of freedom for evaporation, and the evolution stops leaving a droplet disconnected from the black bath. In this case, the decreasing phase of the Page curve ends before reaching zero and stabilizes at a constant value for the entanglement entropy.

Similar phenomena are expected to occur for the evolution of an out-of-equilibrium black funnel in Poincar\'e-AdS. This consists of an AdS black brane that is connected to a Randall-Sundrum brane through a black string funnel. For the horizon on the brane to be hotter than the black brane bath, the funnel must be sufficiently long and thin.\footnote{The flow of this horizon can be given a simple analytic description \cite{Emparan:2013fha}.} Then, the evolution of the GL instability  will lead to a pinch. Where along the black string this pinch occurs will depend on details of the initial state, and it seems possible that pinching at the location of the brane (and therefore the complete evaporation of the black hole) requires fine-tuning. However, a substantial reduction of the size of the horizon on the brane, i.e., incomplete evaporation, seems achievable without fine-tuning.

\paragraph{Evaporation in two-brane systems and  (non-)gravitating baths.} In Sec.~\ref{sec:twin} we have demonstrated different holographic evaporation scenarios in double-brane systems, where the radiation-collecting bath lies in a second brane. Such scenarios have been considered to scrutinize some of the assumptions in the studies of the Page curve \cite{Geng:2020fxl,Geng:2021hlu} (see also \cite{Geng:2021iyq,Miao:2022mdx,Miao:2023unv}). In particular, concerns have been raised that the RT surfaces used in the calculations of \eqref{qes} are anchored in a non-gravitating bath. This could seem a reasonable approximation to the geometry away from the black hole, where gravity is indeed weaker. However, as emphasized in \cite{Geng:2021hlu}, if this region were gravitating, even if weakly, then the anchoring point should be free and subject to extremization, and island surfaces would seemingly disappear. 

We note here that this issue is ameliorated in the large $D$ limit. The large $D$ expansion suppresses the gravitational potential in a very extreme way away from the near-horizon region of the black hole, which has an extent $\propto 1/D$---this is why we took the radial scaling in \eqref{limit} to blow up this region to finite size. The falloff goes as $\sim r^{-D}$, so the spacetime becomes flat (or empty AdS) at any finite distance outside the black hole horizon. Then, this setup makes the toy models of two dimensions exact: radiation will still propagate to the empty region, and we can calculate its entropy using the empty space formula. We may indeed anchor the HRT surface on the brane outside the near-horizon region, and the large $D$ limit will suppress its fluctuations\footnote{Observe that, since in our models the evaporation occurs in a finite time as $D\to\infty$, there does not appear any issue with orders of limits.}. The effect is related to the strong localization of entanglement when $D\to\infty$, as observed in \cite{Colin-Ellerin:2019vst,Giataganas:2021jbj}.  This large-$D$ suppression of the gravitational fluctuations away from the horizon is also present in Randall-Sundrum scenarios, which possibly permits the study of the problem in situations where the graviton on the brane is massless.


\section{Outlook}
\label{sec:outlook}

By explicitly solving the bulk dynamics, we have demonstrated several scenarios for the classical holographic description of evaporating black holes. A crucial feature is that the brane black hole must be connected through the bulk via a funnel-like horizon to a colder horizon that acts as a radiation-collecting bath. An isolated black droplet is stable and does not evaporate. We have only shown a sample of evolutions that exhibit holographic evaporation neatly. Still, more complex behaviors are easily found which can be interpreted by judiciously applying the notions introduced above.

From the dual CFT viewpoint, the nature of the connecting funnel is peculiar: there exist funnels that are dual to states that have zero holographic energy density and pressure, but which nevertheless entangle the $\ord{N^2}$ degrees of freedom of the CFT at different energies. This entanglement allows fast (classical $\sim N^2$) energy transfers from higher to lower temperatures, and thus enables the possibility of the evaporation of a hot black hole into a colder bath at leading large $N$ order. But these states of the CFT are unstable if there is a large enough disparity between the energy scales that are connected, and the evolution of this instability can lead to an intriguing severing of the entanglement between the degrees of freedom at different energies. 

At the end of the day, what our study here and in \cite{Emparan:2021ewh} reveals is that the very rich and distinctive kind of physics that holographic CFTs (as opposed to weakly interacting theories) exhibit when coupled to black holes, is always dual to Gregory-Laflamme-type dynamics of bulk horizons with different length scales along them. The most extreme instance occurs when the horizon pinches to zero. When this happens on the brane, it is the holographic end of complete Hawking evaporation. If instead, it occurs in the bulk, the breakdown of the entangling link between the black hole and the bath brings the Hawking evaporation of the brane black hole to a sudden end before it is completed. 

We emphasize that the main qualitative aspects of our conclusions are not dependent on using the large $D$ approximation. 
This method is successful largely because the emission of bulk gravitational waves is suppressed, but we are not interested in this radiation. The effective theory describes the dynamics of the horizon, which is what we are after, and the main properties of these horizons remain qualitatively similar as $D$ is increased. Therefore, we expect that our simulations capture the basic features of holographic evaporation setups in $D\geq 5$.\footnote{The most important $D$-dependent effect is the existence of a critical dimension for the stability of non-uniform black strings \cite{Sorkin:2004qq}, which in the present context has been identified in \cite{Licht:2022rke}. This mostly adds intricacy of detail in some parameter regions, which here we do not aim to study.}

The large $D$ approach has its limitations, the most significant one being the absence of a sharp distinction between a black hole connected to another system (which can be the asymptotic region or another black hole) by a very thin funnel, and a black hole that is actually disconnected. When $r_0<1$ we can apply sound judgment to discern the situations: we expect that at finite $D$ thin funnels pinch off and split the horizon---either in the bulk or on the brane. However, when $r_0>1$ the funnels are stable and the evolution may stop at them. Another limitation is that we cannot combine in the same system large and small AdS black holes, but this has not been an obstacle to finding a large class of configurations with interesting dynamical evolution.

This framework to model holographic evaporation can be taken further. One may readily chart in more detail the variety of possible evolutions and endpoints that we briefly examined in Sec.~\ref{subsec:remarks}. With some more work, one may also perform the explicit numerical calculations of the Page curve. It is also possible to incorporate the effects of charge, and perhaps rotation, affording qualitatively new possibilities. 

Finally, it would be desirable to have a better understanding, from the CFT viewpoint, of the perplexing properties of black funnels. This should throw light on the intriguing phenomena exhibited by strongly coupled, large-$N$ quantum fields interacting with black holes.

\section*{Acknowledgments}

We gladly acknowledge discussions with Brianna Grado-White, Tom Hartman, Dominik Neuenfeld, David Mateos, Marco Meineri, Mukund Rangamani, Mikel Sanchez-Garitaona-ndia, Jorge Santos, Martín Sasieta, Christoph Uhlemann, and Toby Wiseman. 
RE is especially grateful to Nemanja Kaloper and Takahiro Tanaka for many conversations on holographic evaporation over the years. Work supported by ERC Advanced Grant GravBHs-692951, MICINN grant PID2019-105614GB-C22, AGAUR grant 2017-SGR 754, and State Research Agency of MICINN through the ``Unit of Excellence María de Maeztu 2020-2023” award to the Institute of Cosmos Sciences (CEX2019-000918-M). 
RL acknowledges financial support provided by Next Generation EU through a University of Barcelona Margarita Salas grant from the Spanish Ministry of Universities under the {\it Plan de Recuperaci\'on, Transformaci\'on y Resiliencia} and by Generalitat Valenciana / Fons Social Europeu through APOSTD 2022 post-doctoral grant CIAPOS/2021/150. Work supported by Spanish Agencia Estatal de Investigaci\'on (Grant PID2021-125485NB-C21) funded by MCIN/AEI/10.13039/501100011033 and ERDF A way of making Europe, and the Generalitat Valenciana (Grant PROMETEO/2019/071). 
RS is supported by JSPS KAKENHI Grant Number JP18K13541 and partly by Toyota Technological Institute Fund for Research Promotion A. 
The work of MT is supported by the European Research Council (ERC) under the European Union’s Horizon 2020 research and innovation program (grant agreement No 852386).

\appendix

\section{Diagrams}\label{app:diagrams}

\begin{figure}[h]
        \centering
         \includegraphics[width=1\textwidth]{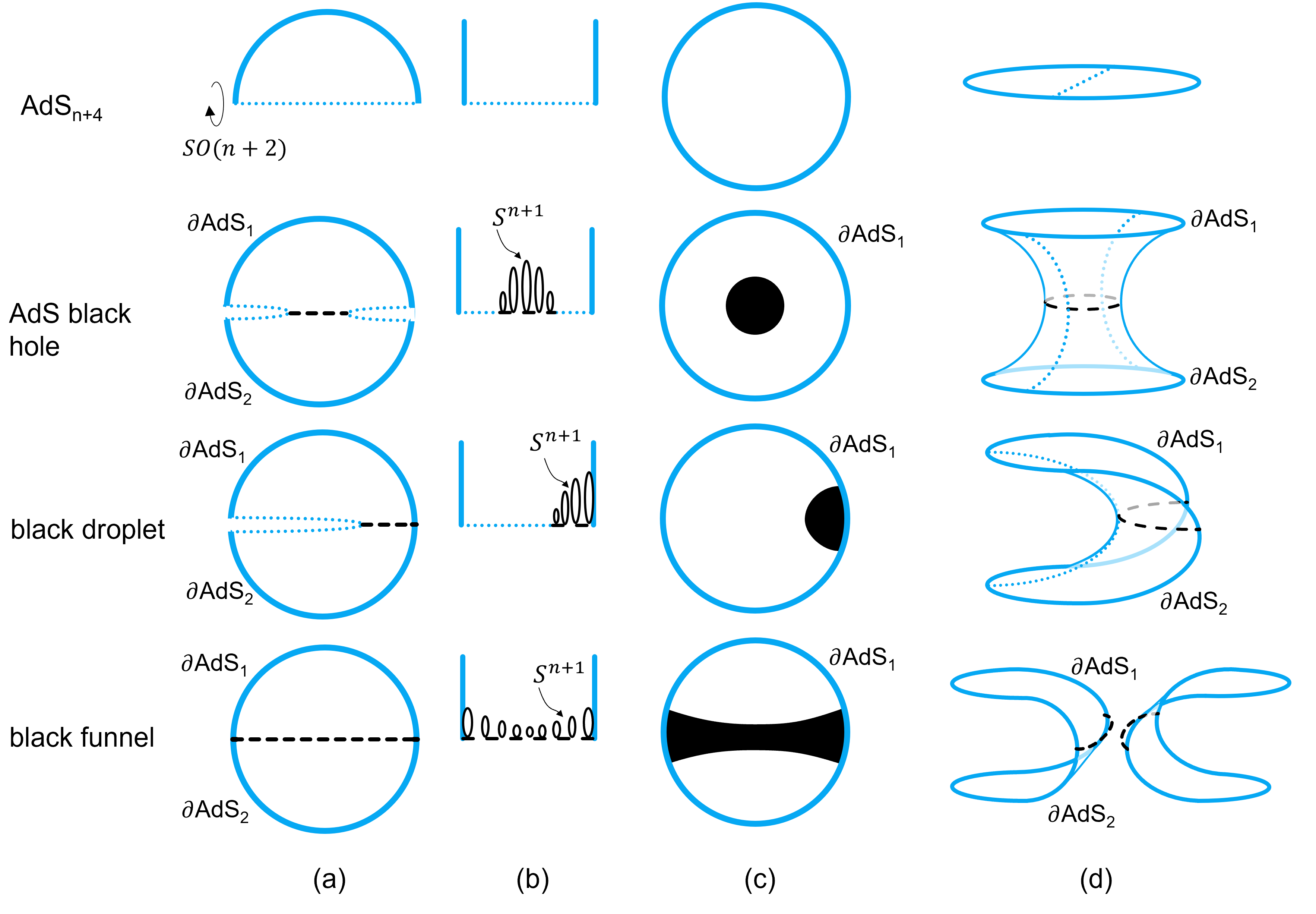}
        \caption{\small Different representations of constant-time slices of the Penrose diagram for configurations in this article (without the brane included). (a) All boundary components are presented. Each point is a sphere $S^{n+1}$ generated by the action of $SO(n+2)$ rotations around the symmetry axis (dotted blue), where the spheres shrink to zero. Dashed black lines are Einstein-Rosen bridges that connect different asymptotic regions. (b) Shape of the horizon, as shown by the size of the $S^{n+1}$ spheres along the ER bridge. (c) Combination of the information in (a) and (b) but showing only a single boundary component. (d) Two antipodal copies of the geometries in (a), with bending chosen for illustrative purposes. When the $SO(n+2)$ rotation is applied, all the horizons, including the black funnel, are connected as in (b).}
        \label{fig:diagrams}
\end{figure}
In figure~\ref{fig:diagrams} we provide a guide for understanding the different representations of black hole configurations in AdS that we employ in this article. All the diagrams represent a constant-time, Cauchy slice of the Penrose diagram of the AdS spacetime in $D=n+4$ dimensions. For simplicity, we do not introduce the brane, but there is no difficulty in adding it.  

We exhibit some representatives cases: empty AdS space, a black hole at the center of AdS (such as the Schwarzschild-AdS solution), a black droplet, and an AdS black string, \eqref{adsbstring}, also known as the uniform black funnel. In the left column, (a), we show all the boundary components of the geometry (black hole spacetimes have two disconnected boundaries, $\partial$AdS$_1$ and $\partial$AdS$_2$). Each point in the diagram is a sphere $S^{n+1}$ generated by the action of the $SO(n+2)$ symmetry group in the geometries \eqref{ansatz} (the extension to the planar and hyperbolic cases is easy). These spheres shrink to zero at the dotted blue line that represents the axis of rotation. The dashed black line is the Einstein-Rosen bridge, where the $S^{n+1}$ have minimal non-zero size, providing a passage between the two asymptotic regions. The (a) diagrams do not show the shape of the horizon. For this purpose, we present the middle column, (b), which depicts the size of the $S^{n+1}$: along the Einstein-Rosen bridge, and along the symmetry axis (where their size is zero). The next two diagrams on the right combine the information: in (c) this is done at the expense of showing only a single asymptotic region; in (d) two antipodal copies of the geometries in (a) (hence with all asymptotic regions) are presented, bent for illustrative purposes.

When the horizon reaches the boundary, it splits it into disconnected components. Note that diagram (a) for the black funnel is the same as for Rindler-AdS, which has indeed been used as a model for a funnel \cite{Chen:2020uac, Chen:2020hmv}, although in that case, strictly, the boundary is connected.

\bibliography{refs}{}
\bibliographystyle{JHEP}

\end{document}